\documentclass[cernpreprint, atlasdraft=false, texlive=2023, UKenglish, texmf, orcidlogo]{atlasdoc}
 
\usepackage{atlaspackage}
\usepackage{atlasbiblatex}
 
\usepackage{atlasphysics}

\graphicspath{{logos/}{figures/}}
 
\usepackage{ANA-TOPQ-2021-29-PAPER-defs}

\AtlasTitle{Measurement of the charge asymmetry in top-quark pair production in association with a photon with the ATLAS experiment}
 
\AtlasVersion{2.3}
 
\AtlasAbstract{
A measurement of the charge asymmetry in top-quark pair ($t\bar{t}$) production in association with a photon is presented. The measurement is performed in the single-lepton $t\bar{t}$ decay channel using proton--proton collision data collected with the ATLAS detector at the Large Hadron Collider at CERN at a centre-of-mass-energy of 13~TeV during the years 2015--2018, corresponding to an integrated luminosity of 139~fb$^{-1}$. The charge asymmetry is obtained from the distribution of the difference of the absolute rapidities of the top quark and antiquark using a profile likelihood unfolding approach. It is measured to be $A_\text{C}=-0.003 \pm 0.029$ in agreement with the Standard Model expectation.
}
 
\AtlasRefCode{TOPQ-2021-29}
\PreprintIdNumber{CERN-EP-2022-260}

\AtlasJournalRef{\PLB 843 (2023) 137848}
\AtlasDOI{10.1016/j.physletb.2023.137848}

\hypersetup{pdftitle={ATLAS document},pdfauthor={The ATLAS Collaboration}}
 
\begin{document}
 
\maketitle

\section{Introduction}
\label{sec:intro}
 
The top quark is the heaviest known elementary particle and the only quark that decays before hadronisation, which allows direct access to its properties in production and decay. Measurements of top-quark properties, predicted by the Standard Model (SM), provide important input to test theoretical calculations and have the potential to reveal deviations from the SM predictions. One of the relevant properties is related to the slight difference between the rapidity distributions of top quarks and top antiquarks produced in pairs (\ttbar). This asymmetry, referred to as charge asymmetry, is defined in proton--proton collisions as follows~\cite{Aguilar-Saavedra:2014vta, Aguilar-Saavedra:2014kpa, Jung:2011zv, Diener:2009ee}:
\begin{equation*}
\Ac = \frac{N(|y_{t}|>|y_{\bar{t}}|) - N(|y_{t}|<|y_{\bar{t}}|)}{N(|y_{t}|>|y_{\bar{t}}|) + N(|y_{t}|<|y_{\bar{t}}|)},
\end{equation*}
 
where $N$ is the number of events and $y_{t}$ ($y_{\bar{t}}$) the rapidity of the top quark (top antiquark). The production of \ttbar events is predicted to be symmetric under the exchange of top quark and
antiquark, i.e.\ $\Ac= 0$, at leading-order (LO) accuracy in perturbative QCD. However, at next-to-leading order (NLO), quark--antiquark-initiated \ttbar production is asymmetric in the top-quark rapidity distribution, owing to interference between processes with initial- and final-state gluon emission and between the Born diagram and the box diagram at $O(\alphas^4)$. The total asymmetry from the sum of all effects is expected to be positive~\cite{PhysRevD.86.034026}.
 
Previous measurements of the asymmetry in \ttbar production by the ATLAS and CMS Collaborations at centre-of-mass energies ($\sqrt{s}$) of 7, 8 and 13~\TeV~\cite{TOPQ-2012-17,TOPQ-2014-16,CMS-TOP-11-030,CMS-TOP-12-033,TOPQ-2013-09,TOPQ-2015-08,CMS-TOP-12-010,CMS-TOP-15-009,TOPQ-2015-10,CMS-TOP-13-013,CMS-TOP-17-014} agree with the SM expectation. The most recent measurement of the inclusive and differential charge asymmetry at 13~\TeV\ by the ATLAS Collaboration~\cite{arXiv:2208.12095} reported evidence for a non-zero asymmetry in \ttbar production (measured as $\Ac=0.0068\pm0.0015$ and in agreement with the SM prediction~\cite{PhysRevD.98.014003,arXiv:2208.12095}). While \Ac at the LHC corresponds to a central--forward asymmetry, the \ttbar production asymmetry manifests itself as a forward--backward asymmetry at the Tevatron. Early measurements of this asymmetry showed deviations from NLO QCD predictions, particularly at large values of the \ttbar invariant mass~\cite{Aaltonen:2011kc}. However, more recent results by the CDF and D0 Collaborations~\cite{Abazov:2015fna,Aaltonen:2012it,Abazov:2014cca} are compatible with the improved SM predictions including NLO electroweak (EW) and higher-order QCD corrections~\cite{PhysRevLett.115.052001,PhysRevD.91.071502}.
 
\begin{figure}[ht]
\centering
\includegraphics[width=0.25\textwidth]{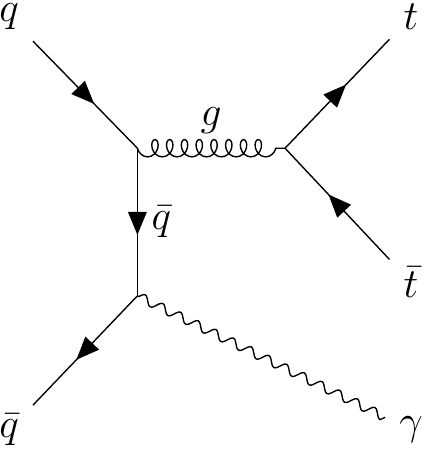}
\quad \quad
\includegraphics[width=0.30\textwidth]{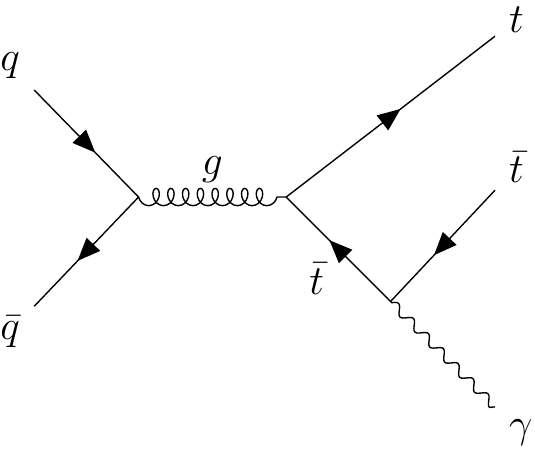}
\caption{Example Feynman diagrams of \tty production contributing to the charge asymmetry.}
\label{fig:intro_diag_tty}
\end{figure}
 
The \ttbar charge asymmetry is diluted at the LHC owing to the large fraction of gluon--gluon-initiated \ttbar events, which are symmetric under the exchange of the top quark and antiquark. However, it is enhanced in other topologies where the fraction of quark--antiquark-initiated production is larger, such as in associated production of \ttbar with a photon (\tty)~\cite{Aguilar-Saavedra:2014vta, Aguilar-Saavedra:2014kpa}. Interference effects among QCD diagrams at NLO, similar to those in \ttbar events, are predicted in \tty production. However, the dominant contribution to the asymmetry in \tty arises from interference between QED initial-state radiation, Figure~\ref{fig:intro_diag_tty}(left), and final-state radiation, Figure~\ref{fig:intro_diag_tty}(right), which yields a larger asymmetry of negative sign. In addition, other QCD--EW higher-order contributions can have a sizeable effect on the observed asymmetry~\cite{Pagani:2021}. The overall asymmetry in \tty at $\sqrt{s}= 13$~\TeV\ is expected to have a negative value, of 1\%--2\% depending on the phase space, according to SM predictions~\cite{Schulze:2019ttyAc, Pagani:2021} and can be modified by \enquote{beyond-the-SM} contributions. Contributions from an $s$-channel colour octet or a $Z'$ boson would, for instance, result in a smaller \Ac absolute value~\cite{Aguilar-Saavedra:2014vta}.
 
These sources of asymmetry are only present in the \tty events where the photon is radiated from an initial-state parton or one of the top quarks (hereafter referred to as \textit{\tty production}). The asymmetry is diluted by \tty events where the photon arises from any of the charged decay products of the \ttbar system (\textit{\tty decay} in the following). Therefore, only \tty production events are considered as signal in this analysis.
 
This \paperornote presents the first measurement of the charge asymmetry of the top-quark pairs in \tty production in \ttbar single-lepton final states, which have one high-\pt lepton and at least four jets, two of which arise from $b$-quarks. It is performed using the full 139~\ifb\ data set recorded with the ATLAS detector between 2015 and 2018 at $\sqrt{s}=13~\TeV$, referred to as \RunTwo. In order to extract the asymmetry, the top quarks are reconstructed using a kinematic likelihood fit. The separation between signal and background processes is enhanced using a neural network (NN) approach. The output distribution of the NN is used to define two regions, one enriched in background events and one in signal events. The \Ac value is determined by means of a maximum-likelihood fit to the distribution of the difference of absolute rapidities of the top quark and antiquark. This is also referred to as \enquote{maximum-likelihood unfolding}.

The \paperornote is organised as follows. The ATLAS detector is briefly introduced in Section~\ref{sec:detector}. The simulation of signal and background processes is summarised in Section~\ref{sec:modelling}. The event reconstruction,  selection, and estimation of the background processes are presented in Sections~\ref{sec:selection} and~\ref{sec:background}. The systematic uncertainties are described in Section~\ref{sec:systematics}. The analysis strategy is discussed in Section~\ref{sec:analysis}, followed by the result in Section~\ref{sec:result}. Finally, a summary is given in Section~\ref{sec:conclusion}.
 
\section{ATLAS detector}
\label{sec:detector}
\newcommand{\AtlasCoordFootnote}{
ATLAS uses a right-handed coordinate system with its origin at the nominal interaction point (IP) in the centre
of the detector and the $z$-axis along the beam pipe. The $x$-axis points from the IP to the centre of the
LHC ring, and the $y$-axis points upwards. Cylindrical coordinates $(r,\phi)$ are used in the transverse plane,
$\phi$ being the azimuthal angle around the $z$-axis. The pseudorapidity is defined in terms of the polar angle
$\theta$ as $\eta = -\ln \tan(\theta/2)$, and the rapidity as $y = (1/2)[(E+p_z)/(E-p_z)]$.
Angular distance is measured in units of $\Delta R \equiv \sqrt{(\Delta\eta)^{2} + (\Delta\phi)^{2}}$.}
 
The ATLAS~\cite{PERF-2007-01,ATLAS-TDR-19,Abbott:2018ikt} detector is a multipurpose detector with a
forward--backward symmetric cylindrical geometry with respect to the LHC beam axis.\footnote{\AtlasCoordFootnote}
The innermost layers consist of tracking detectors in the pseudorapidity range
$|\eta|<2.5$. This inner detector (ID) is surrounded by a thin superconducting solenoid
that provides a \SI{2}{\tesla} axial magnetic field. It is enclosed by the electromagnetic and
hadronic calorimeters, which cover $|\eta|<4.9$. The outermost layers of ATLAS consist of an external
muon spectrometer within  $|\eta|<2.7$, incorporating three large toroidal magnetic assemblies with eight coils each.
The field integral of the toroids ranges between 2.0 and 6.0$\,$Tm for most of the acceptance.
The muon spectrometer includes precision tracking chambers and fast detectors for triggering.
A two-level trigger system~\cite{TRIG-2016-01} reduces the recorded event rate to an average of \SI{1}{\kilo\hertz}. An extensive software suite~\cite{ATL-SOFT-PUB-2021-001} is used in data simulation, in the reconstruction and analysis of real and simulated data, in detector operations, and in the trigger and data acquisition systems of the experiment.
 
\section{Simulation of signal and background processes}
\label{sec:modelling}
Monte Carlo (MC) event generators were used to estimate the contributions from the expected signal and background processes. The response of the ATLAS detector was simulated~\cite{SOFT-2010-01} with \GEANT~\cite{Agostinelli:2002hh}.
A fast simulation (\textsc{AtlFast-II}), which relies on a parameterisation of the calorimeter response~\cite{ATL-PHYS-PUB-2010-013}, was used in samples employed to estimate uncertainties related to the \ttbar and \tWy modelling. The additional \pp~collisions in the same or neighbouring bunch crossings, referred to
as pile-up, were generated with \PYTHIA[8.186]~\cite{Sjostrand:2007gs} using a set of tuned parameters called the A3 tune~\cite{ATL-PHYS-PUB-2016-017} and the \NNPDF[2.3lo] parton distribution function (PDF) set~\cite{Pumplin:2002vw}.
 
The signal \tty production events were simulated with \MGNLO[2.7.3]~\cite{Alwall:2014hca} as a $2\to 3$ process at NLO accuracy in QCD. The interference effects between initial-state and final-state photon radiation were considered in this process. The final-state top quarks in the \tty production sample are on-shell and were decayed at LO using \MADSPIN~\cite{Frixione:2007zp,Artoisenet:2012st} to preserve spin correlations. 
 
The background \tty decay events, where the photon arises from any of the decay products of the top quarks or one of the on-shell top quarks, were simulated with the same version of \MGNLO but at LO precision as a $2 \to 2$ process followed by the decay of the top quarks, also simulated at LO precision. Both samples were generated using the \NNPDF[3.0nlo]~\cite{Ball:2014uwa} PDF set and interfaced to \PYTHIA[8.240]~\cite{Sjostrand:2014zea}, which used the A14 tune~\cite{ATL-PHYS-PUB-2014-021} and the \NNPDF[2.3lo] PDF set to model the parton shower, hadronisation, fragmentation and underlying event. The renormalisation and factorisation scales were set to $0.5\times \sum_i \sqrt{m^2_i+p^2_{\text{T},i}}$, where $m_i$ and $p_{\text{T},i}$ are the masses and transverse momenta of the particles generated from the matrix element (ME) calculation. Photons are required to have $\pT > 15~\GeV$ and to be isolated according to a smooth-cone hadronic isolation criterion with $\delta_0=0.1$, $\epsilon_{\gamma}=0.1$ and $n=2$, defined in Ref.~\cite{Frixione:1998jh}, which avoids infrared divergences. The top-quark mass in the \tty sample and all other samples involving top quarks was set to \SI{172.5}{\gev} and the decays of bottom and charm hadrons were simulated using the \EVTGEN[1.6.0] program~\cite{Lange:2001uf}.
 
The \tty production sample is normalised to the NLO cross section given by the MC simulation, while the normalisation of the \tty decay sample is corrected by a NLO/LO inclusive $K$-factor of 1.5. This $K$-factor was derived by comparing the normalisation of the sum of the NLO \tty production sample and the LO \tty decay sample with the normalisation of a LO inclusive $2 \to 7$ \tty sample corrected with the $K$-factor obtained in Ref.~\cite{TOPQ-2017-14} using the calculation described in Ref.~\cite{Melnikov:2011ta}.
 
The \ttbar events were simulated at NLO accuracy in QCD using \POWHEGBOX[v2]~\cite{Nason:2004rx,Frixione:2007vw,Alioli:2010xd} and the \NNPDF[3.0nlo] PDF set. The parton shower was  generated with \PYTHIA[8.230] using the A14 tune~\cite{ATL-PHYS-PUB-2012-003}. The \ttbar events are normalised to a cross-section value calculated with the \TOPpp[2.0] program at next-to-next-to-leading order (NNLO) in perturbative QCD, including soft-gluon resummation to next-to-next-to-leading-logarithm order (see Ref.~\cite{Czakon:2011xx:mod} and references therein).
 
The \tWy events were generated at LO accuracy with the \MGNLO[2.7.3] generator in the five-flavour scheme. To simulate this process, two complementary samples were generated; one as a $2\rightarrow3$ process assuming a stable top quark and the other as a $2 \to 2$ process, where the photon is radiated from any other charged final-state particle. To avoid infrared divergences, the photon was required to have $\pT > 15~\GeV$ and $|\eta| < 5.0$ and to be separated by $\Delta R>0.2$ from any parton. Both samples make use  of the \NNPDF[2.3lo] PDF set and were interfaced to \PYTHIA[8.212] for parton showering using the A14 tune.
 
Single-top-quark $s$- and $t$-channel production and inclusive $tW$ production were simulated at ME level at NLO in QCD with \POWHEGBOX[v2] and the \NNPDF[2.3lo] PDF set. The event generator was interfaced to \PYTHIA[8.230], which used the A14 tune. The events are normalised to the NNLO cross section~\cite{Kidonakis:2010tc,Kidonakis:2010ux,Kidonakis:2011wy}.

Events with $W\gamma$~and $Z\gamma$~final states (with additional jets) were simulated with \SHERPA[2.2.8]~\cite{Bothmann:2019yzt} at NLO in QCD using the \NNPDF[3.0nnlo] PDF set. The samples are normalised to the cross sections given by the MC simulation. The \SHERPA generator performs all steps of the event generation, from the hard process to the observable particles.
Events with inclusive $W$- and $Z$-boson production in association with additional jets were simulated with \SHERPA[2.2.1]~\cite{Bothmann:2019yzt,Hoeche:2009rj} at NLO in QCD.
The \NNPDF[3.0nlo] PDF set was used together with a dedicated tune provided by the \SHERPA authors.
The samples are normalised to the NNLO cross section in QCD~\cite{STDM-2015-03}.
 
Diboson processes, $\mathit{WW}$, $\mathit{WZ}$ and $\mathit{ZZ}$, were generated with \SHERPA[2.2.2] (leptonic decays) and~2.2.1 (non-leptonic final states) at LO in QCD. The \NNPDF[3.0nnlo] PDF set was used with a dedicated tune provided by the \SHERPA authors.
The samples are normalised to NLO cross sections in QCD~\cite{Campbell:1999ah}.
 
Events with a \ttbar pair and an associated $W$~or $Z$~boson ($t\bar{t}V$) were simulated at NLO at the ME level with \MGNLO using the \NNPDF[3.0nlo] PDF set. The ME generator was interfaced to \PYTHIA[8.210], for which the A14 tune was used in conjunction with the \NNPDF[2.3lo] PDF set. The samples are normalised to NLO in QCD and electroweak theory~\cite{deFlorian:2016spz}.

A procedure was applied to remove the overlap between the samples in which events were generated at ME level without explicitly including a photon in the final state and the dedicated samples where photons were included in the ME-level event-generation step (\tty and \Wty final states, as well as $W\gamma$ and $Z\gamma$ final states with additional jets). Events in the inclusive samples are discarded if they contain a parton-level photon that fulfils $\pT(\gamma)>\SI{15}{\gev}$ and $\Delta R(\gamma, \ell) > 0.2$, where $\pT(\gamma)$ is the transverse momentum of the photon and $\Delta R(\gamma, \ell)$ is the angular distance between the photon and any charged lepton.
 
Corrections to the pile-up profile, the trigger, reconstruction and selection efficiencies, and the energy scales and resolutions, are applied to the MC simulation samples to improve the description of the data.

\section{Event reconstruction and selection}
\label{sec:selection}
 
The analysis uses a data set that passes stringent quality requirements and corresponds to an integrated luminosity of 139~\ifb{} collected with the ATLAS detector during \RunTwo of the LHC. Events are required to have at least one primary vertex reconstructed from at least two associated tracks, and only events where at least one single-electron~\cite{TRIG-2018-05} or single-muon~\cite{TRIG-2018-01} trigger was fired are selected.
 
Muons are reconstructed by combining a track in the muon spectrometer with a track in the ID system.
The reconstruction, identification and calibration methods are described in Ref.~\cite{MUON-2018-03}. The muon track is also required to originate from the primary collision vertex. Muons are required to be isolated according to measurements of nearby track \pT and calorimeter energy.
Only muons with calibrated $\pT > \SI{25}{\GeV}$ and $|\eta| < 2.5$ and passing `medium' quality requirements are considered.
 
Electrons are reconstructed from energy deposits in the electromagnetic calorimeter (ECAL) associated with reconstructed tracks in the ID system. The origin of the electron track also has to be compatible with the primary vertex. Electrons are identified with a combined likelihood technique~\cite{EGAM-2018-01} using a `tight' working point, and are required to be isolated  according to measurements of nearby calorimeter energy and track \pT. Electrons are calibrated with the method described in Ref.~\cite{EGAM-2018-01} and are selected if they fulfil $\pt>\SI{25}{\GeV}$ and $|\eta_{\mathrm{clus}}| < 2.47$, excluding the ECAL barrel/endcap transition region $1.37 < |\eta_{\mathrm{clus}}| < 1.52$, where $\eta_{\mathrm{clus}}$ refers to the pseudorapidity of the calorimeter energy cluster associated with the electron.

Photons are reconstructed from energy deposits in the central region of the ECAL~\cite{EGAM-2018-01}. Photons are required to fulfil tight identification and isolation requirements. The latter is defined as $\left. \ET^{\mathrm{iso}} \right|_{\Delta R < 0.4} < 0.022 \cdot \ET(\gamma) + \SI{2.45}{\GeV}$ in conjunction with $\left. \pT^{\mathrm{iso}} \right|_{\Delta R < 0.2} < 0.05 \cdot \ET(\gamma)$, where $\ET^{\mathrm{iso}}$ refers to the calorimeter isolation within a cone of size $\Delta R = 0.4$ around the direction of the photon candidate and $\pT^{\mathrm{iso}}$ is the track isolation within a $\Delta R = 0.2$ cone~\cite{PERF-2017-02}.
Photons are required to have transverse energy $\ET > \SI{20}{\GeV}$ and $|\eta_{\mathrm{clus}}| < 2.37$, excluding the calorimeter transition region. They are separated into two categories, one where the cluster is not matched to any reconstructed track in the ID system (unconverted photons) and the other where the cluster is matched to one or two reconstructed tracks that are consistent with originating from a photon conversion and, in addition, a conversion vertex can be found (converted photons).

Jets are reconstructed using the anti-$k_{t}$ algorithm~\cite{Cacciari:2008gp} in the \Fastjet implementation~\cite{Fastjet} with a distance parameter $R=0.4$. Their reconstruction is performed on particle-flow objects~\cite{PERF-2015-09}. The jet energy scale and jet energy resolution are calibrated using an energy- and $\eta$-dependent calibration scheme based on simulation with in situ corrections obtained from data~\cite{PERF-2016-04}. Only jets with $\pT > \SI{25}{\GeV}$ and $|\eta| < 2.5$ are considered in the analysis.
Jets with a large contribution from pile-up vertices are identified with the \emph{jet vertex tagger} (JVT)~\cite{ATLAS-CONF-2014-018} and rejected.
 
Jets arising from \bquark hadronisation, referred to as \bjets, are identified using the DL1r \btag algorithm~\cite{ATLAS:2022qxm}, which is based on an artificial deep neural network combining information from other algorithms using track impact parameters and secondary vertices, and a multi-vertex reconstruction algorithm. The flavour-tagging efficiency for \bjets, as well as for $c$-jets and light-flavour jets, is calibrated as described in Ref.~\cite{FTAG-2018-01}. The working point used to select the \bjets corresponds to a selection efficiency of 77\% in simulated \ttbar events.
 
The magnitude of the reconstructed missing transverse momentum (\met)~\cite{ATLAS-CONF-2018-023,PERF-2016-07} is calculated from the negative vector sum of the \pT of all calibrated physics objects and the remaining unclustered energy, also called the \emph{soft term}. This term is estimated from low-\pt tracks associated with the primary vertex but not with any reconstructed object.
 
An overlap removal procedure is implemented to avoid the reconstruction of the same energy clusters or tracks as different objects. Electron candidates that share their track with a muon candidate are removed and jets within a $\Delta R = 0.2$ cone around any of the remaining electrons are excluded. If the distance between an electron and any remaining jet is $\Delta R < 0.4$, the electron is subsequently removed. In the next step, muon candidates within $\Delta R = 0.4$ of a jet are removed if the jet has more than two associated tracks, otherwise the jet is discarded. In the final step, photons within a $\Delta R = 0.4$ cone around any remaining electron or muon are excluded and then jets within a $\Delta R = 0.4$ cone around any remaining photon are removed.
 
Events are selected if they have exactly one electron or one muon that is matched to the corresponding trigger-level object. The \pt thresholds for the leptons are $25~\GeV$ in 2015 data, $27~\GeV$ in 2016 data, and $28~\GeV$ in 2017 and 2018 data, which are at least 1~\GeV\ above the \pt thresholds of the single-lepton triggers. This is done to avoid differences due to the calibration of the objects used in the trigger logic, and objects used in the physics analysis.
Only events containing exactly one reconstructed photon fulfilling the condition $\Delta R(\ell,\gamma)>0.4$ are considered in the measurement. Additionally, events where the invariant mass of the electron--photon system is within $5~\GeV$ of the $Z$-boson mass are rejected. The event is also required to have at least four jets and at least one of them must be \btagged.
 
The kinematic properties, in particular the rapidities of the top quark and antiquark, are determined by means of a constrained kinematic fitting algorithm, KLFitter~\cite{KLFitter:2014}, based on a maximum-likelihood approach applied to the four-momenta of the selected lepton and up to five leading jets (\pT-ordered) and \met, representing the transverse momentum of the neutrino. The likelihood is constructed as the product of transfer functions that relate the energies of the reconstructed objects and parton-level objects and Breit--Wigner distributions that match the selected objects to the $W$ bosons and the top quarks. The fit is constrained to reconstruct two $W$ bosons, each with a mass of 80.4~\GeV. In addition, the reconstructed masses of the top quark and antiquark are constrained to 172.5~\GeV. The combination of jets that gives the highest likelihood is selected and the jets are used to reconstruct the hadronically and leptonically decaying top quark or antiquark. In the case of leptonic top-quark decay, the invariant mass of the lepton--neutrino system is constrained to be the $W$-boson mass and a quadratic equation for the neutrino's longitudinal momentum is obtained. For real solutions of the equation, the solution that results in the top-quark mass closer to 172.5~\GeV\ is chosen, while in the case of complex solutions, the real part of the solution is considered. The fraction of top quarks that are reconstructed within $\Delta R=1.0$ of the parton-level top quarks is about 64\% for hadronically decaying top quarks and 72\% for leptonically decaying top quarks.
 
\section{Background estimation}
\label{sec:background}
Each background process is assigned to one of several categories depending on the origin of the reconstructed photon, whether it is a prompt photon or whether another object mimics a photon signature. The estimation of background events from different sources closely follows the methods employed in Ref.~\cite{TOPQ-2017-14}.
 
After the event selection, the largest background contribution is that of \tty decay events (about 30\% of the total number of events). The \emph{\catprompty background} category, which contains any other type of background process with a prompt photon, constitutes about 15\% of the selected events. Both contributions are estimated using MC simulation. Agreement between data and simulation of the \catprompty background was validated in dedicated regions. A validation region enriched in the $Z\gamma$ process is defined by selecting events with exactly one photon, two same-flavour opposite-sign leptons, fewer than four jets and no $b$-tagged jets. Events in the $W\gamma$ validation region fulfil the same lepton and photon requirements as in the signal region. The simulation of $W\gamma$ is known to underestimate the number of events with heavy-flavour jets in data. Thus, to define a region orthogonal to the signal region while selecting events with heavy-flavour content, the events in the validation region are additionally required to have fewer than four jets and at least one must pass the $b$-tagging working point with a selection efficiency of 85\% but fail the one with 70\% efficiency in simulated \ttbar events. The expected fractions of $Z\gamma$ and $W\gamma$ events in the corresponding control regions are about 95\% and 50\%, respectively.
 
Another significant contribution arises from processes with an electron mimicking a photon signature in the detector, referred to as \emph{e-fake}. This background contribution amounts to 16\% of the total number of selected events and it is estimated from data by applying a tag-and-probe method to $Z\rightarrow e^+e^-$ events~\cite{PERF-2017-02}. Two control regions are defined in order to determine the \efake rate in data and simulation: one region contains events with an electron--positron pair and the other, enriched in \efake events, contains events with an electron and a photon satisfying the object selection criteria described in Section~\ref{sec:selection}. Additionally, the invariant mass of the pair of objects is required to be in the range [40, 140]~\GeV, and their angular separation in $\phi$ must exceed 2.62~rad to suppress the contributions from events where the photon is radiated from the electron. Background contributions not originating from $Z$-boson events are subtracted in data with a fit of the invariant mass
distribution. The $Z \rightarrow e^+e^-\gamma$ contribution with prompt photons where one of the electrons is not reconstructed or identified is subtracted using simulation. The \efake rate is obtained as the ratio of the event yield after background subtraction in the e-fake-enriched control region to the yield in the electron--positron control region. The calculated ratio is binned in the \pt (three bins) and $|\eta|$ (four bins) of the photon in the $e\gamma$ events and either the electron or the positron in the $e^+e^-$ events, selected randomly to avoid biasing the selection, and separately for converted and unconverted photons. Scale factors are calculated to correct the e-fake background estimate in the signal region, based on a comparison between the e-fake photon rates obtained using either data or simulation. The systematic uncertainties in the scale factors account for possible mismodelling of the signal and background processes in the fit. The values of the scale factors vary from 0.8 to 1.4 with uncertainties between 5\% and 20\%.

The background contribution from events where the photon signature arises from hadronic energy depositions in the ECAL or from hadron decays such as $\pi^0 \to \gamma\gamma$, generically referred to as \emph{\cathfake}, constitutes about 7\% of the events. The \cathfake background is estimated from data by using the so-called ABCD method. Three orthogonal regions enriched with \cathfake photon events are defined by inverting the photon isolation selection and the requirements on four variables related to the shower shape in the first layer of the ECAL, which are part of the photon tight identification criteria. They are chosen because of their small correlation with the photon isolation and their power to discriminate between prompt and \cathfake photons. Events are selected for regions A and B if their photon fails at least two out of four identification requirements, while satisfying all other identification criteria, and pass or fail the isolation requirements, respectively. Region C contains events where the photons fail the isolation requirements but satisfy the tight identification criteria. Additionally, the sum of the \pt of all tracks within $\Delta R=0.2$ of the photon is required to be larger than 3~\GeV\ to further suppress the prompt-photon contribution in regions B and C. The \cathfake background contribution in the signal region is measured as the product of the numbers of events in regions A and C divided by the number of events in region B. The estimate is corrected for the correlation between the criteria, which is obtained using MC simulations. The scale factors are obtained separately for converted and unconverted photons and as a function of the photon \pt (two bins) and $|\eta|$ (four bins). The considered sources of systematic uncertainty in the \cathfake background contribution include the modelling of the \ttbar process, which contributes about 90\% of the \cathfake events, the shower shapes, and the normalisation uncertainties of the background processes. The scale factors range from 0.6 to 1.5, with uncertainties of 30\%--60\%.
 
The contribution from events with a non-prompt or misidentified lepton, referred to as \emph{lepton fake}, is obtained using the data-driven approach referred to as the matrix method~\cite{ATLAS:2022swp}. Events are separated into two categories that are based on tighter or looser lepton identification and isolation requirements, and thus enriched in events with real leptons or non-prompt/fake leptons, respectively. The contribution in the signal region is estimated from the data events passing loose lepton selection requirements, corrected by a weight that depends on the real- and fake-lepton efficiencies obtained from the two event categories described above. The efficiencies are parameterised as a function of the lepton kinematic properties. The uncertainties are estimated by using different parameterisations and tighter control regions, and by including the normalisation uncertainty of the prompt-lepton background. This background contribution amounts to around 1\% of all selected events, dominated by events with a misidentified electron.
 
\section{Systematic uncertainties}
\label{sec:systematics}
The precision of the obtained asymmetry \Ac is affected by several sources of systematic uncertainty, arising from detector effects or theoretical assumptions, as well as uncertainties due to the limited number of events in the MC simulations. The different sources of systematic uncertainty, discussed in the following, affect the event yields, the distribution shape of the observable of interest, or both. The sources of systematic uncertainty affecting the shape of the distribution typically have a larger impact on the precision of the result because global normalisation uncertainties cancel out in the ratio of event yields that defines \Ac.
 
The experimental systematic effects include uncertainties in the integrated luminosity and the simulation of pile-up events, as well as effects related to the reconstruction and identification of the physics objects in the analysis.
 
The uncertainty in the total integrated luminosity is estimated to be 1.7\%~\cite{ATLAS-CONF-2019-021}, using the LUCID-2 detector \cite{LUCID2} for the primary luminosity measurements.  The uncertainty associated with the modelling of pile-up is determined by varying the pile-up reweighting in the simulation within its uncertainties.
 
The photon and lepton identification and isolation efficiencies, momentum scale and resolution~\cite{PERF-2017-03,PERF-2015-10}, and lepton trigger efficiencies in simulation are corrected using scale factors to better describe the corresponding values in data. These corrections, which typically depend on \pT and $\eta$, are varied within their uncertainties to estimate the corresponding systematic uncertainty.
 
The jet energy scale (JES) uncertainty is derived from a combination of simulations, test-beam data and in situ measurements~\cite{PERF-2016-04}. Contributions from jet-flavour composition, $\eta$-intercalibration, punch-through, single-particle response, calorimeter response to different jet flavours, and pile-up are also taken into account, yielding a total of 30 uncorrelated JES uncertainty subcomponents, of which 29 are non-zero in a given event depending on the type of simulation used. The jet energy resolution in simulation is smeared by its corresponding uncertainty~\cite{ATL-PHYS-PUB-2015-015} split into eight uncorrelated sources. The uncertainty associated with the JVT discriminant for pile-up jet rejection is obtained by varying the efficiency correction factors.
 
The uncertainties in the $b$-jet tagging calibration are determined separately for $b$-jets, $c$-jets and light-flavour jets~\cite{PERF-2016-05,FTAG-2020-08,ATLAS-CONF-2018-006}. For each jet category, the uncertainties are decomposed into several uncorrelated components.
 
The uncertainty in \MET arises from the propagation of the energy scales and resolutions of photons, leptons and jets, and the modelling of its soft term~\cite{PERF-2016-07}.
 
The signal and background modelling uncertainties include those owing to the choice of QCD scales, parton shower, amount of QCD initial-state radiation (ISR), and PDF set.
 
The effect of the QCD scale uncertainty for each of the \tty, \Wty and \ttbar processes is evaluated independently by separately halving and doubling the renormalisation and factorisation scales relative to the default scale choice. The uncertainty from the parton shower and hadronisation for those processes is estimated by comparing the nominal simulated samples interfaced with \PYTHIA[8] with alternative samples interfaced to \HERWIG[7]~\cite{Bahr:2008pv,Bellm:2015jjp}. Uncertainties due to the value of \alphas used in the ISR parton shower modelling are estimated by comparing the nominal \tty, \Wty and \ttbar simulations with alternative samples simulated with higher or lower radiation parameter settings in the A14 tune, controlled by the \textit{var3c} parameter implemented in \PYTHIA[8].
An additional ISR uncertainty is obtained for the \ttbar process by comparing the nominal sample with an additional one with the \hdamp parameter, which controls the \pt of the first additional emission, varied by a factor of two~\cite{ATL-PHYS-PUB-2018-009}. The uncertainty in the PDFs for the signal and background \tty processes is estimated using the 30 PDF variations of the PDF4LHC15 prescription~\cite{Butterworth:2015oua}. The PDF variations are propagated by using alternative generator weights and each of them is considered as a separate nuisance parameter in the fit.
 
For the e-fake, h-fake, and lepton-fake background contributions, the total uncertainties associated with the corrections obtained using data are propagated to the final result. A normalisation uncertainty of 20\% is assigned to the \tty decay process, based on the estimated uncertainty in the NLO $K$-factor~\cite{TOPQ-2017-14}, and a 50\% normalisation uncertainty is assigned to $W\gamma$, based on the differences between data and simulation observed in the dedicated control region, and to the minor background processes contributing to the prompt-photon category, i.e.\ single top quark, \ttV, diboson, and $Z\gamma$.

\section{Signal discrimination}
\label{sec:analysis}
A multivariate analysis using a neural network is performed to further separate the \tty signal from the background processes. The NN is fully connected and consists of three hidden layers. The first two layers consist of 96 nodes and are followed by a batch normalisation layer. The third layer has 16 nodes. The hidden layers use a parametric ReLU activation function, while the output node uses a sigmoid activation function. The training is performed with Keras~\cite{chollet2015keras} with the TensorFlow~\cite{tensorflow2015-whitepaper} backend with binary cross-entropy as a loss function. The overall events are split into a training and validation set (85\%) and a testing set (15\%). The first set of events is used in a 5-fold cross-validation: Events are split randomly into 5 folds, the model is trained on 4 folds and one fold is used for validation of the NN configuration. This procedure is repeated 5 times. The event weights are applied to the events in the training, testing and validation samples. The NN uses a total of 21 variables related to the kinematic properties of individual objects, such as the photon \pt and $\eta$, event shape variables (e.g.\ \met and the scalar sum of the \pt of the jets in the event), the number of $b$-tagged jets, the pseudo-continuous binned $b$-tagging discriminant~\cite{FTAG-2018-01}, the photon conversion type, and invariant masses and angular separations of different objects in the event (e.g.\ the invariant mass and $\Delta R$ of the lepton or the photon and the closest $b$-tagged jet). Example variables from among those with the most discriminating power are shown in Figure~\ref{fig:prefit_variables_1}. The MC simulation describes the data within the uncertainties. The largest contributions to the MC uncertainty band are the normalisation uncertainties associated with the backgrounds with prompt photons.
 
\begin{figure}[ht]
\centering
\includegraphics[width=0.3\textwidth]{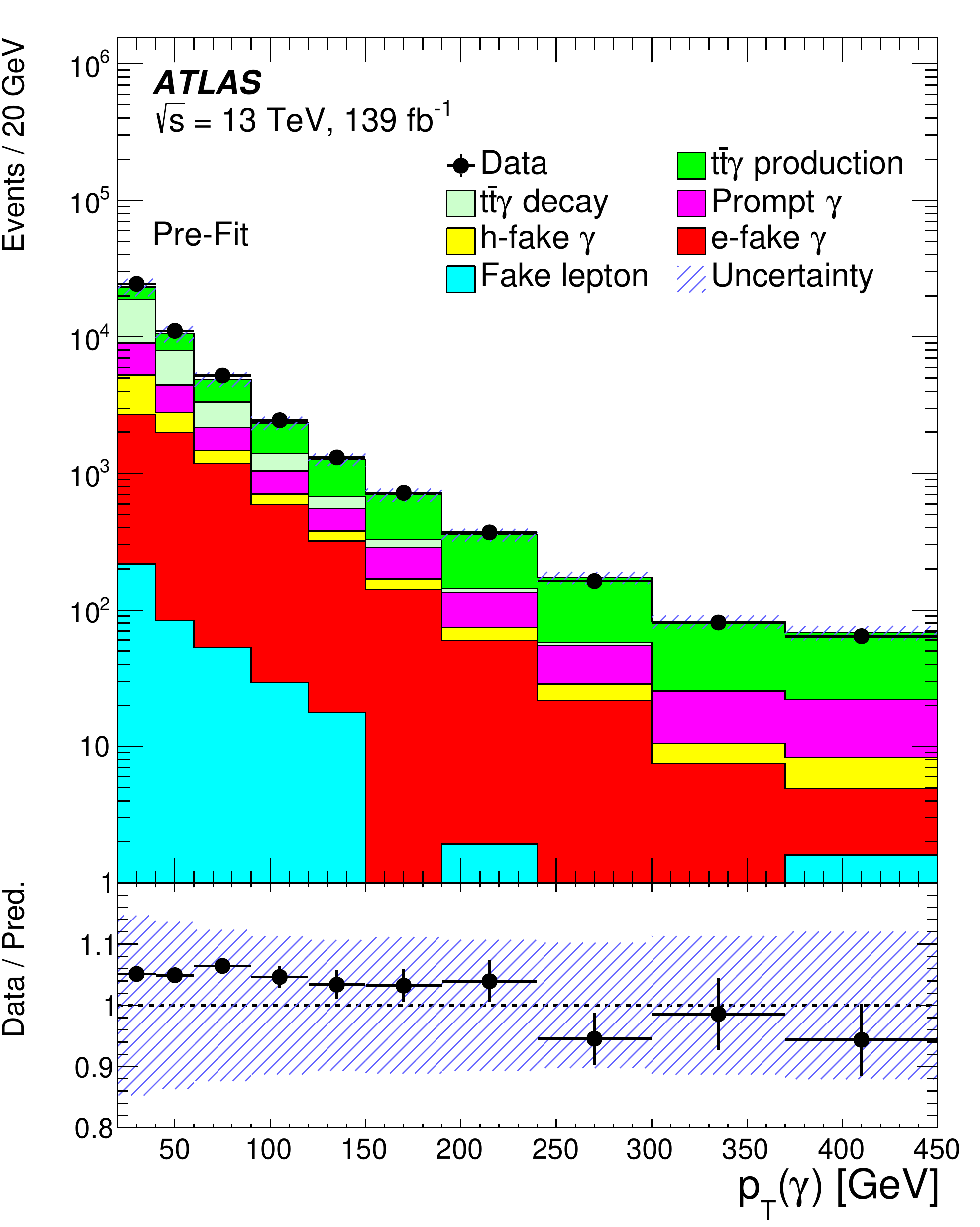}
\includegraphics[width=0.3\textwidth]{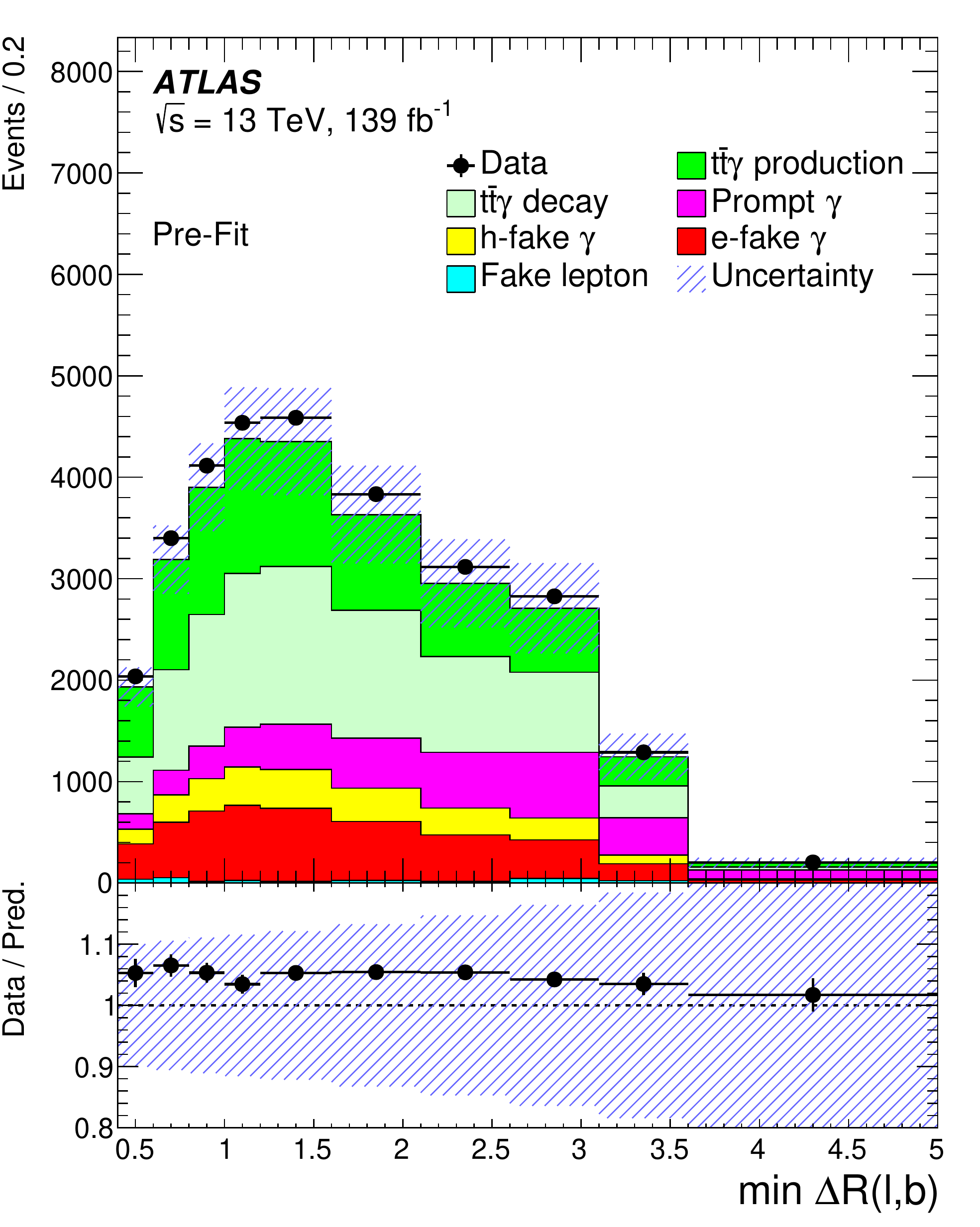}
\includegraphics[width=0.3\textwidth]{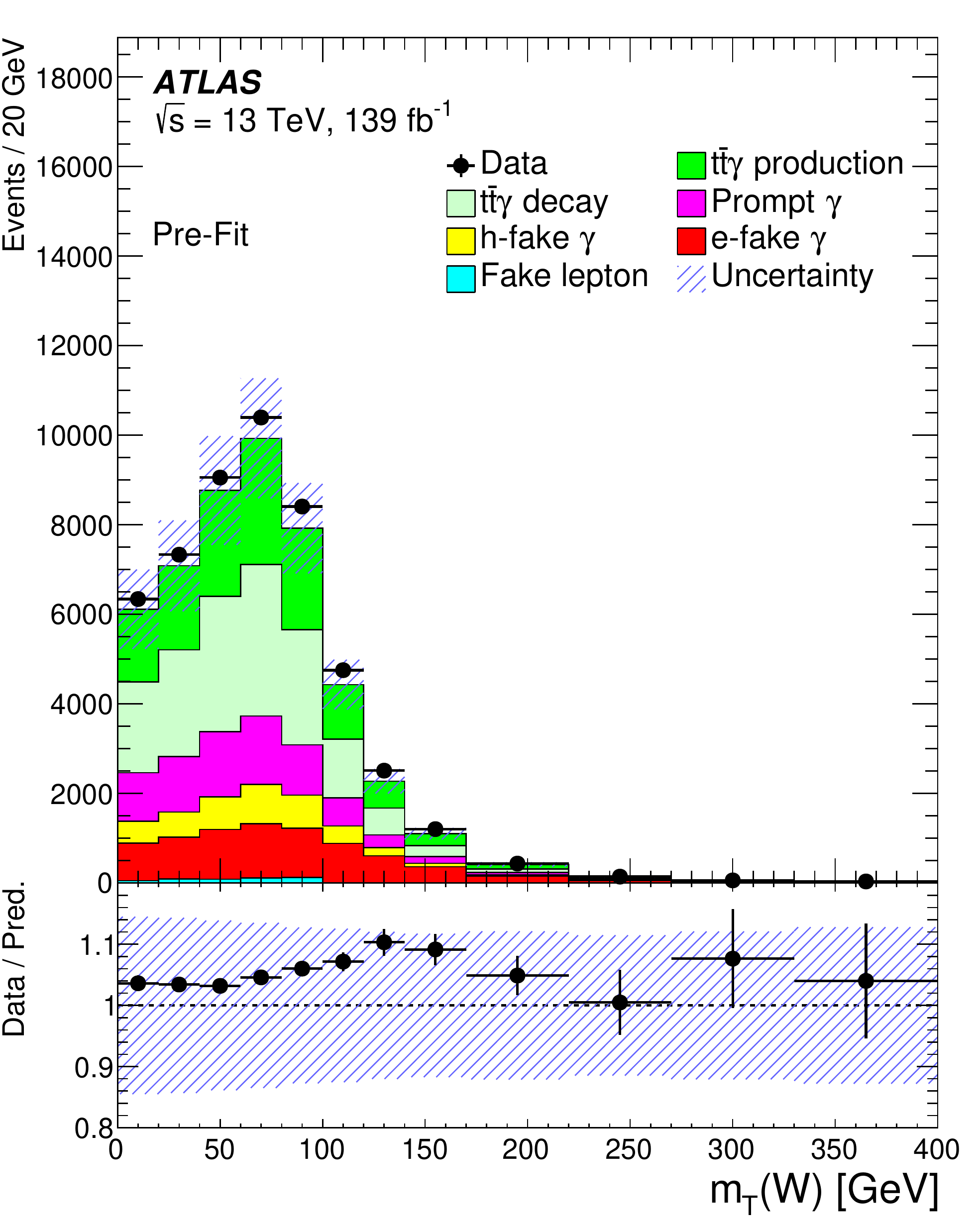}
\caption{Distributions of photon \pt (left), angular separation of the lepton and the closest $b$-tagged jet (middle), and transverse mass of the leptonically decaying $W$ boson (right) before the fit. The uncertainty band includes all experimental and modelling systematic uncertainties (cf.\ Section~\ref{sec:systematics}) added in quadrature. Overflow events are included in the last bin of each distribution. The lower part of the plot shows the ratio of the data to the prediction.}
\label{fig:prefit_variables_1}
\end{figure}

The resulting NN discriminant output, $O_{\mathrm{NN}}$, shown in Figure~\ref{fig:prefit_variables_NN}, is used to divide the events into a background-enriched region and a signal-enriched region, defined by $O_{\mathrm{NN}} < 0.6$ and $O_{\mathrm{NN}} \geq 0.6$, respectively. This threshold was optimised to give the smallest expected uncertainty in \Ac. The observed and expected signal and background event yields in the two regions are summarised in Table~\ref{tab:prefityields}. The slight underestimate of the data by the SM prediction, observed in Figure~\ref{fig:prefit_variables_1}, is reflected in Figure~\ref{fig:prefit_variables_NN} at large values of $O_{\mathrm{NN}}$ because it is expected to partially come from the normalisation of the \tty production simulation, which is a free parameter in the profile likelihood unfolding described in the following.

\begin{figure}[ht]
\centering
\includegraphics[width=0.45\textwidth]{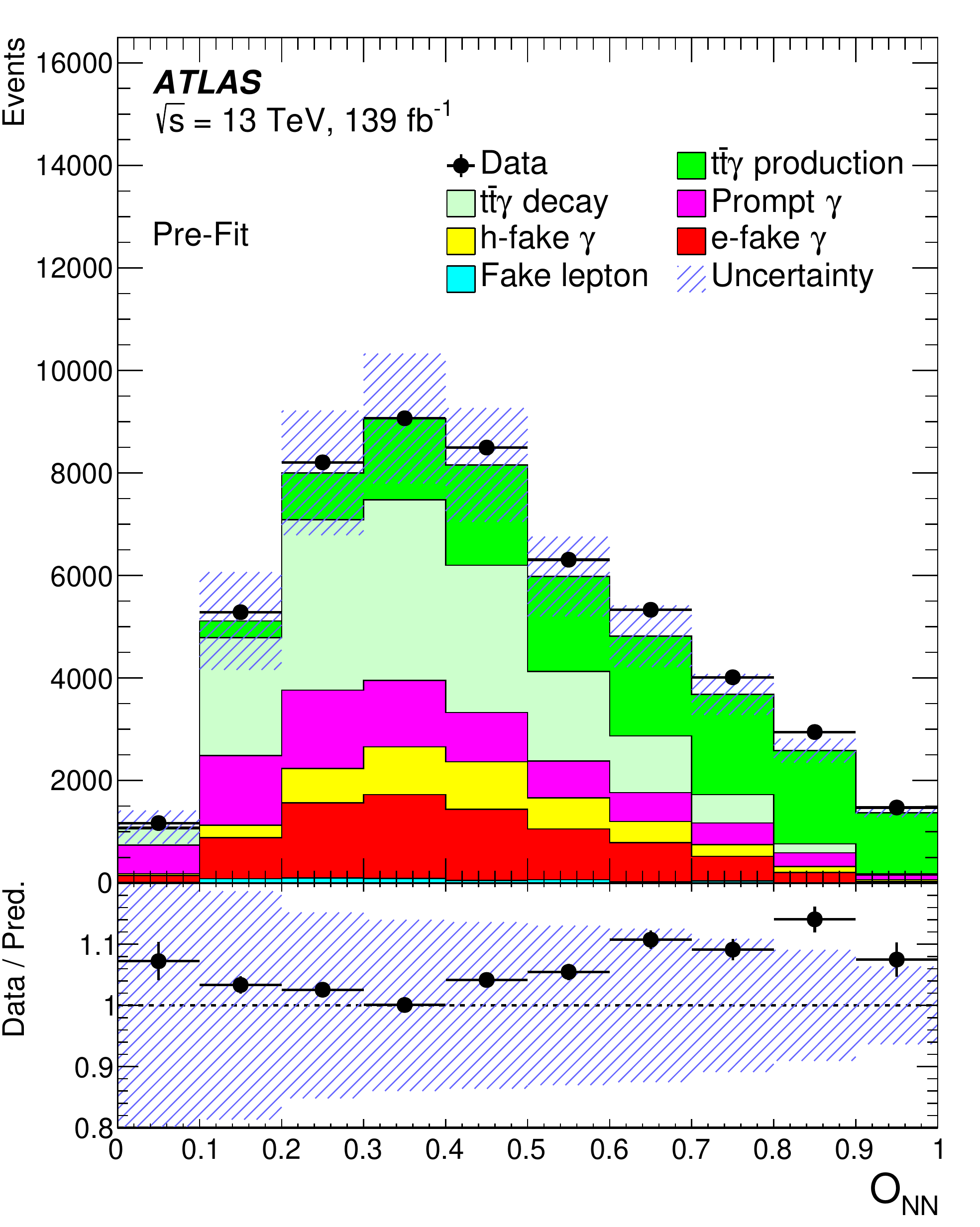}
\caption{Distribution of the NN output discriminant before the fit. The uncertainty band includes all experimental and modelling systematic uncertainties (cf.\ Section~\ref{sec:systematics}) added in quadrature. The lower part of the plot shows the ratio of the data to the prediction.}
\label{fig:prefit_variables_NN}
\end{figure}

\begin{table}
\centering
\caption{Event yields before the profile likelihood unfolding after the full selection in the two regions defined by the NN discriminant value. The quoted uncertainties correspond to all statistical and systematic uncertainties (cf.\ Section~\ref{sec:systematics}) added in quadrature.}
\label{tab:prefityields}
\sisetup{table-format=4.0(3)}
\footnotesize
\begin{tabular}{l|r|r}
\hline
& {$O_{\mathrm{NN}} < 0.6$} & {$O_{\mathrm{NN}} \geq 0.6$}\\
\hline
\tty prod (signal)   	& 6660 	$\pm$  \phantom{0}350 	& 6910 	$\pm$  \phantom{0}340 \\
\tty decay   		& 14\,100 	$\pm$ 3100 			& 1900 	$\pm$  \phantom{0}560 \\
h-fake $\gamma$   	& 3400 	$\pm$ 1400 			& 790 	$\pm$  \phantom{0}360 \\
e-fake $\gamma$   	& 6420 	$\pm$  \phantom{0}860 	& 1480 	$\pm$  \phantom{0}260 \\
Prompt $\gamma$  & 6400 	$\pm$ 2000 			& 1300 	$\pm$  \phantom{0}400 \\
Lepton fake   		& 410 	$\pm$  \phantom{0}110 	& 57 		$\pm$  \phantom{00}35 \\
\hline
Total  & \numRF{37405.9}{3} $\pm$ \numRF{4543.05}{2} & \numRF{12438.5}{3} $\pm$ \numRF{1055.71}{2} \\
 
\hline
Data   & 38\,527\phantom{~$\pm$~0000} & 13\,763\phantom{~$\pm$~0000} \\
 
\hline
\end{tabular}
\end{table}
 
\section{Results}
\label{sec:result}
The value of \Ac is extracted from the $|y_{t}| - |y_{\bar{t}}|$ distribution in a fiducial region defined at particle level. The top quark and antiquark are defined at parton level in the MC simulation after final-state radiation but before decay. The fiducial region at particle level is defined by applying selection requirements similar to those at reconstruction level to the stable particles after the event generation and before the detector simulation. The fiducial phase space is defined by requiring exactly one photon, exactly one electron or muon, and at least four jets, of which at least one must be a $b$-jet, defined as follows. Photons are required to not originate from a hadron decay, to have $\et > \SI{20}{\GeV}$ and $|\eta| < \SI{2.37}{}$, and to be isolated such that the sum of transverse momenta of all charged particles surrounding the photon within $\Delta R \leq 0.2$ must be less than 5\% of its own \pt. Muons and electrons must have $\pt>25$~\GeV\ and $|\eta| < 2.5$, and must not originate from hadron decays. The momenta of nearby photons, within a $\Delta R = 0.1$ cone, are added to the lepton before applying the selection. Jets are clustered with the anti-$k_{t}$ algorithm with a radius parameter of $R=0.4$. All stable particles are considered in the clustering, except for the selected electrons, muons and photons, and the neutrinos originating from the top quarks. Jets are required to have $\pt > 25~\GeV$ and $|\eta| < 2.5$. A particle-level jet is identified as a $b$-jet if a hadron with $\pT > \SI{5}{\GeV}$ containing a $b$-quark is matched to the jet through a ghost-matching method~\cite{Cacciari:2008gn}. Jets within $\Delta R = 0.4$ of lepton or isolated photon candidates are removed.
 
The \Ac value is obtained by means of a simultaneous maximum-likelihood unfolding of the $|y_{t}| - |y_{\bar{t}}|$ distributions in the two regions defined by the NN output discriminant. The efficiency of selecting and reconstructing an event that is generated in the fiducial phase space is about 30\% in the two regions, while the fraction of events that fulfil the selection at reconstruction level but fail the particle-level requirements is about 20\%. The fraction of events that are reconstructed in the $|y_{t}| - |y_{\bar{t}}|$ bin where they were generated is approximately 75\%. The parameters of interest, which float freely in the fit, are the signal strength of the bin $|y_{t}| - |y_{\bar{t}}|>0$ and \Ac, which replaces the signal strength of the other bin, using the following expression: $\Ac = (\mu_{+}T_{+} - \mu_{-}T_{-})/(\mu_{+}T_{+} + \mu_{-}T_{-})$. The signal strength $\mu_{+}$ $(\mu_{-})$ is defined as the ratio of the measured cross section to the expected value given by the SM simulation in the bin $|y_{t}| - |y_{\bar{t}}|>0~(<0)$ at generator level. The variable $T_{+/-}$ represents the number of \tty production events at particle level in the corresponding bin. No regularisation is applied.  The systematic uncertainties are taken into account via nuisance parameters in the likelihood function. They are symmetrised, taking half of the difference between the upward and downward variations as the uncertainty. If both variations have the same sign, the average of the difference between each variation and the nominal value is chosen as the two-sided uncertainty. If only one variation is available (e.g.\ the uncertainty from the parton shower or the PDF uncertainties) the difference from the nominal value is taken as both the upward and downward variations for the corresponding source. In addition, a pruning procedure is implemented in order to remove the smallest systematic uncertainties.
 
In addition to the NLO/LO $K$-factor applied to correct the normalisation of the \tty decay process, the $|y_{t}| - |y_{\bar{t}}|$ template is reweighted to account for the asymmetry in \ttbar production. The weight is obtained at parton level by considering the central value of the prediction for the inclusive \ttbar asymmetry, $\Ac^{\ttbar} = 0.0064^{+0.0005}_{-0.0006}$, calculated at NNLO accuracy in QCD with EW corrections at NLO~\cite{PhysRevD.98.014003,arXiv:2208.12095}. The weight is then propagated to the distribution at reconstruction level. For consistency, the MC templates obtained with the NLO \ttbar samples are also reweighted to match that asymmetry value. To probe the robustness of the method, and in particular to verify that the unfolding procedure does not bias the result towards the asymmetry in the signal MC simulation, the measurement was repeated with pseudodata. Several pseudodata sets were obtained by reweighting the $|y_{t}| - |y_{\bar{t}}|$ distribution of the signal \tty production sample to correspond to different \Ac values and by adding the background contributions. The unfolding procedure was repeated using the nominal simulation. The resulting values of the asymmetry are in agreement with the true \Ac asymmetry of each pseudodata set within the statistical precision and do not indicate any bias.
 
\begin{figure}[ht]
\centering
\includegraphics[width=0.45\textwidth]{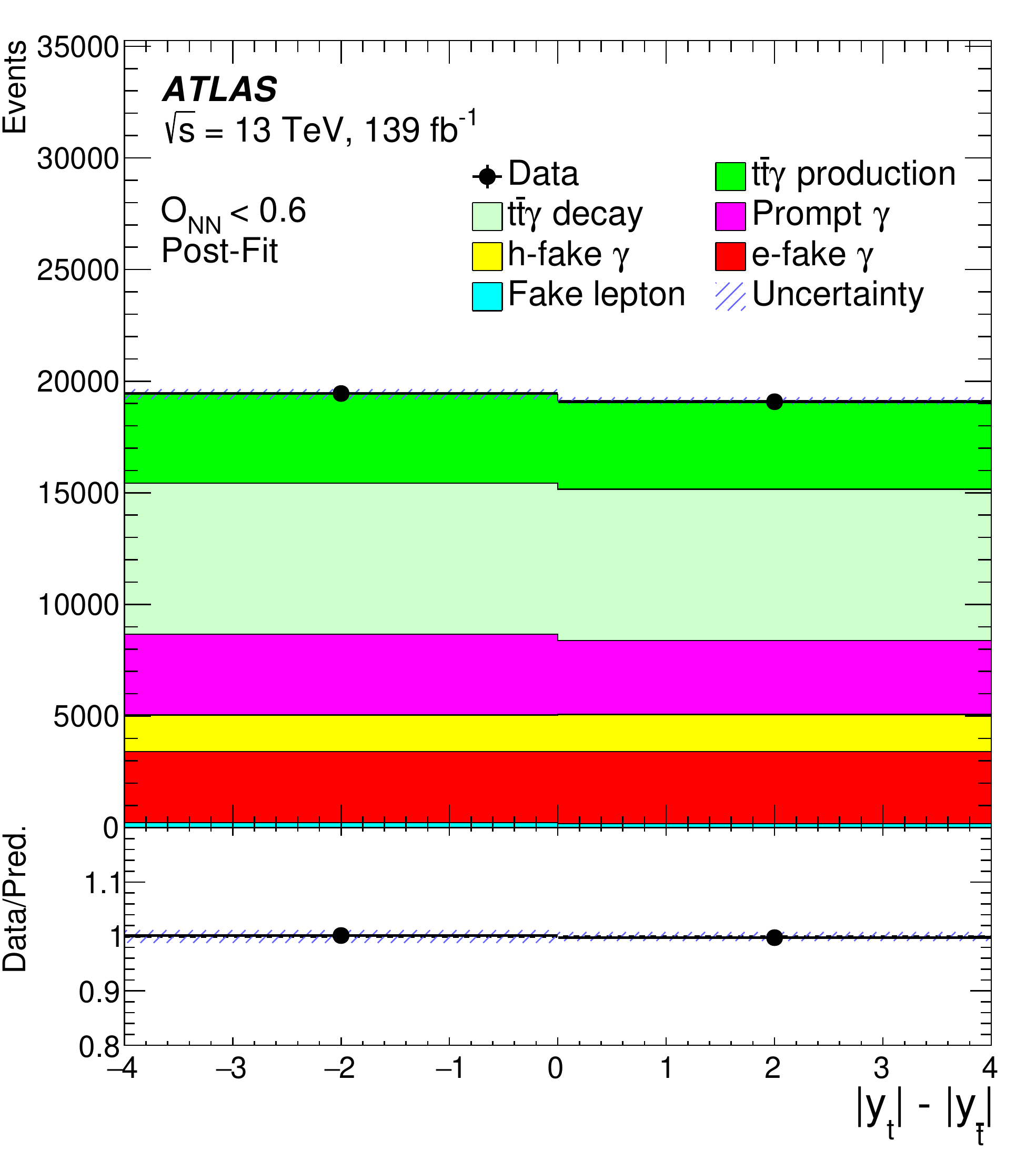}
\includegraphics[width=0.45\textwidth]{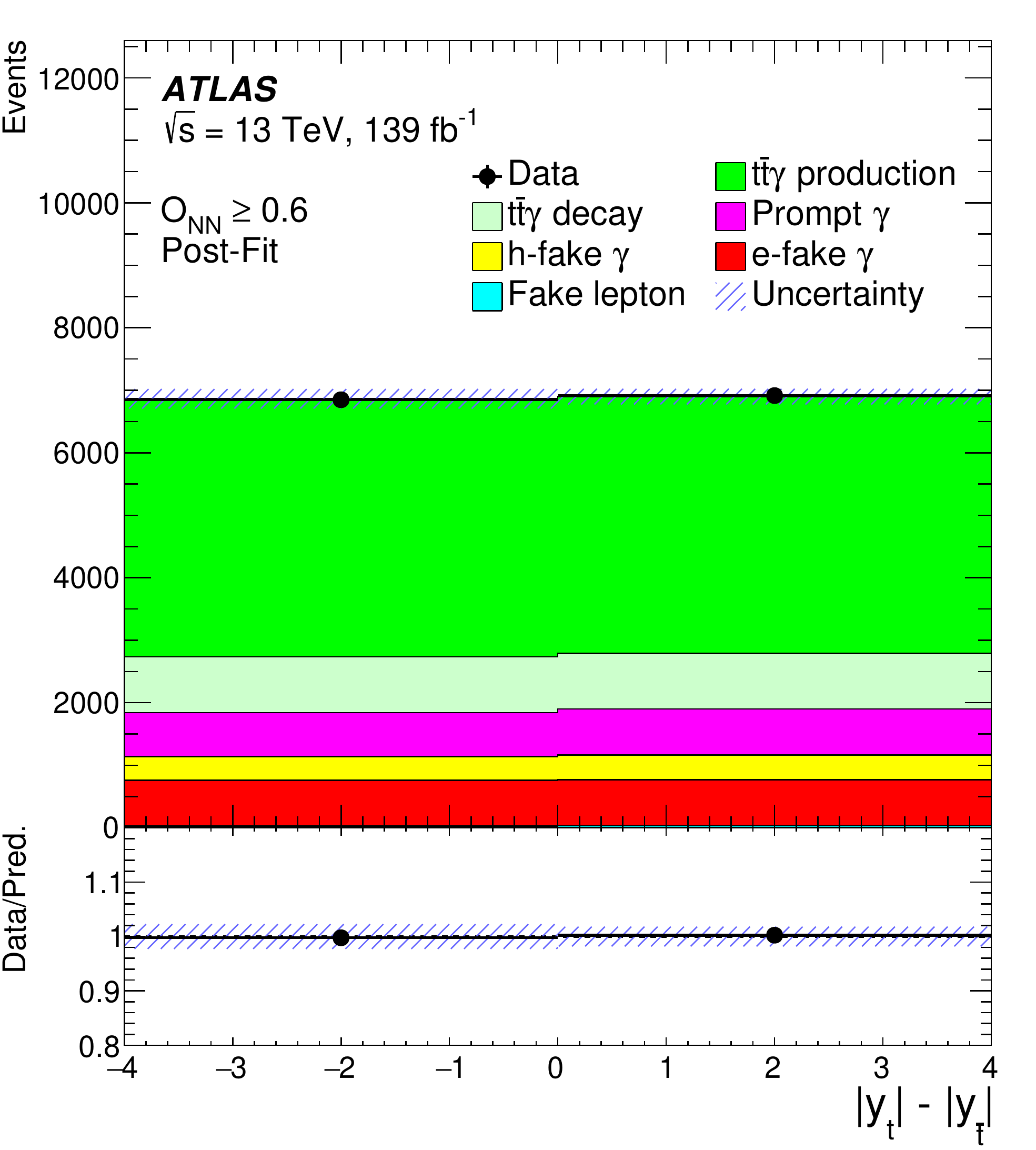}
\caption{The distributions of $|y_{t}| - |y_{\bar{t}}|$ after the fit in the two regions defined by the NN output. Underflow and overflow events are included in corresponding bins of the distributions. The uncertainty band represents the total post-fit uncertainties. Correlations among uncertainties are taken into account as determined in the fit. The lower part of the plot shows the ratio of the data to the prediction.}
\label{fig:dy_postfit}
\end{figure}
 
The post-fit $|y_{t}| - |y_{\bar{t}}|$ distributions in the two regions are shown in Figure~\ref{fig:dy_postfit}. The \Ac of the \tty production process is expected to have a negative sign, while the asymmetry of the background contributions with a top-quark pair, i.e.\ \tty decay and \ttbar, is expected to be positive as discussed in Section~\ref{sec:intro}. Good agreement is observed between the data and the prediction after the fit. As a result of the fit, a few nuisance parameters are slightly constrained: the uncertainties in the normalisation of the \tty decay and $W\gamma$ backgrounds are reduced by 30\% and 15\%, respectively. In all cases, the best-fit values of the nuisance parameters are well within one standard deviation of their initial values.
 
The asymmetry is found to be $\Ac=-0.003 \pm 0.029 = -0.003 \pm 0.024(\mathrm{stat}) \pm 0.017(\mathrm{syst})$, assuming the SM \ttbar charge asymmetry of $\Ac^{\ttbar}=0.0064$~\cite{PhysRevD.98.014003}. The systematic uncertainty is derived from its squared value, calculated as the difference of the squares of the total uncertainty and the statistical uncertainty, obtained from a fit without systematic uncertainties. The \Ac value is compatible with the value obtained from the \MGNLO MC simulation in the same phase space, $\Ac=-0.014\pm0.001(\mathrm{scale})$. The precision of the result is limited by the statistical uncertainty. The impact of the different sources of systematic uncertainty, grouped in categories, is summarised in Table~\ref{tab:syst_impact}. The most relevant sources of systematic uncertainty are the MC statistical uncertainty of the prompt-photon background and the experimental systematic sources related to jets and \MET. The dependence of the measured \tty \Ac on $\Ac^{\ttbar}$ is estimated by repeating the measurement for different values of the \ttbar asymmetry in the range between 0 and $2\times \Ac^{\ttbar}$. The dependence is found to be linear and can be parameterised as $\Ac = -0.57\times \Ac^{\ttbar}+0.0005$.

\begin{table}[htbp]
\centering
\caption{
Summary of the impact of the systematic uncertainties on \Ac grouped into different categories. The quoted uncertainties are obtained by repeating the fit with certain sets of nuisance parameters fixed to their post-fit values, and calculating the squared uncertainties as the difference of the squares of the full-fit and repeated-fit uncertainties. The category \textit{Other experimental} includes uncertainties associated with leptons, pile-up and luminosity.}
\label{tab:syst_impact}
\footnotesize
\begin{tabular}{l S}
\hline
\textbf{Total uncertainty} & 0.029 \\
\hline
\textbf{Statistical uncertainty} & 0.024 \\
\hline
\textbf{MC statistical uncertainties} & \\
\midrule
Background processes & 0.008 \\
\tty production & 0.004 \\
\midrule
\textbf{Modelling uncertainties} & \\
\midrule
\tty production modelling & 0.003 \\
Background modelling & 0.002 \\
Prompt background normalisation & 0.002 \\
\midrule
\textbf{Experimental uncertainties} & \\
\midrule
Jet & 0.009 \\
Fake-lepton background estimate & 0.005 \\
\MET & 0.005 \\
Fake-photon background estimates & 0.003 \\
Photon & 0.001 \\
$b$-tagging & 0.001 \\
Other experimental & 0.004 \\
\hline
\end{tabular}
\end{table}
 
\FloatBarrier
 
\section{Conclusion}
\label{sec:conclusion}
 
This \paperornote presents a measurement of the top-quark pair charge asymmetry in \tty events using \intlumitot of $pp$ collision data at a centre-of-mass energy of 13~\TeV\ collected by the ATLAS experiment at the LHC. The selected events have exactly one photon, one lepton, and at least four jets, of which at least one is $b$-tagged. The inclusive charge asymmetry yields $\Ac=-0.003 \pm 0.029 = -0.003 \pm 0.024(\mathrm{stat}) \pm 0.017(\mathrm{syst})$, which is compatible with the Standard Model prediction within the uncertainties. The precision is limited by the statistical uncertainty.
 
\section*{Acknowledgements}


We thank CERN for the very successful operation of the LHC, as well as the
support staff from our institutions without whom ATLAS could not be
operated efficiently.
 
We acknowledge the support of
ANPCyT, Argentina;
YerPhI, Armenia;
ARC, Australia;
BMWFW and FWF, Austria;
ANAS, Azerbaijan;
CNPq and FAPESP, Brazil;
NSERC, NRC and CFI, Canada;
CERN;
ANID, Chile;
CAS, MOST and NSFC, China;
Minciencias, Colombia;
MEYS CR, Czech Republic;
DNRF and DNSRC, Denmark;
IN2P3-CNRS and CEA-DRF/IRFU, France;
SRNSFG, Georgia;
BMBF, HGF and MPG, Germany;
GSRI, Greece;
RGC and Hong Kong SAR, China;
ISF and Benoziyo Center, Israel;
INFN, Italy;
MEXT and JSPS, Japan;
CNRST, Morocco;
NWO, Netherlands;
RCN, Norway;
MEiN, Poland;
FCT, Portugal;
MNE/IFA, Romania;
MESTD, Serbia;
MSSR, Slovakia;
ARRS and MIZ\v{S}, Slovenia;
DSI/NRF, South Africa;
MICINN, Spain;
SRC and Wallenberg Foundation, Sweden;
SERI, SNSF and Cantons of Bern and Geneva, Switzerland;
MOST, Taiwan;
TENMAK, T\"urkiye;
STFC, United Kingdom;
DOE and NSF, United States of America.
In addition, individual groups and members have received support from
BCKDF, CANARIE, Compute Canada and CRC, Canada;
PRIMUS 21/SCI/017 and UNCE SCI/013, Czech Republic;
COST, ERC, ERDF, Horizon 2020 and Marie Sk{\l}odowska-Curie Actions, European Union;
Investissements d'Avenir Labex, Investissements d'Avenir Idex and ANR, France;
DFG and AvH Foundation, Germany;
Herakleitos, Thales and Aristeia programmes co-financed by EU-ESF and the Greek NSRF, Greece;
BSF-NSF and MINERVA, Israel;
Norwegian Financial Mechanism 2014-2021, Norway;
NCN and NAWA, Poland;
La Caixa Banking Foundation, CERCA Programme Generalitat de Catalunya and PROMETEO and GenT Programmes Generalitat Valenciana, Spain;
G\"{o}ran Gustafssons Stiftelse, Sweden;
The Royal Society and Leverhulme Trust, United Kingdom.
 
The crucial computing support from all WLCG partners is acknowledged gratefully, in particular from CERN, the ATLAS Tier-1 facilities at TRIUMF (Canada), NDGF (Denmark, Norway, Sweden), CC-IN2P3 (France), KIT/GridKA (Germany), INFN-CNAF (Italy), NL-T1 (Netherlands), PIC (Spain), ASGC (Taiwan), RAL (UK) and BNL (USA), the Tier-2 facilities worldwide and large non-WLCG resource providers. Major contributors of computing resources are listed in Ref.~\cite{ATL-SOFT-PUB-2021-003}.


\printbibliography

@article{KLFitter:2014,
      author         = "Johannes Erdmann and Stefan Guindon and Kevin Kroeninger and Boris Lemmer and Olaf Nackenhorst and Arnulf Quadt and Philipp Stolte",
      title          = "{A likelihood-based reconstruction algorithm for top-quark pairs and the KLFitter framework}",
      journal        = "Nucl. Instrum. Meth. A",
      volume         = "748",
      year           = "2014",
      pages          = "18",
      doi            = "10.1016/j.nima.2014.02.029",
      eprint         = "1312.5595",
      archivePrefix  = "arXiv",
      primaryClass   = "hep-ex",
}

@article{Pagani:2021,
    author = "Pagani, Davide and Shao, Hua-Sheng and Tsinikos, Ioannis and Zaro, Marco",
    title = "{Automated EW corrections with isolated photons: t$ \overline{t} $\ensuremath{\gamma}, t$ \overline{t} $\ensuremath{\gamma}\ensuremath{\gamma} and t\ensuremath{\gamma}j as case studies}",
    eprint = "2106.02059",
    archivePrefix = "arXiv",
    primaryClass = "hep-ph",
    reportNumber = "DESY 21-083, DESY-21-083, LU-TP 21-19, TIF-UNIMI-2021-7",
    doi = "10.1007/JHEP09(2021)155",
    journal = "JHEP",
    volume = "09",
    pages = "155",
    year = "2021"
}

@article{Schulze:2019ttyAc,
      author         = "Jonas Bergner and Markus Schulze",
      title          = "{The top quark charge asymmetry in \tty production at the LHC}",
      journal        = "Eur. Phys. J. C",
      volume         = "79",
      year           = "2019",
      pages          = "189",
      doi            = "10.1140/epjc/s10052-019-6707-6",
      eprint         = "1812.10535",
      archivePrefix  = "arXiv",
      primaryClass   = "hep-ph",
      reportNumber   = "HU-EP-18/42",
}

@article{Cacciari:2008gn,
      author         = "Cacciari, Matteo and Salam, Gavin P. and Soyez, Gregory",
      title          = "{The catchment area of jets}",
      journal        = "JHEP",
      volume         = "04",
      year           = "2008",
      pages          = "005",
      doi            = "10.1088/1126-6708/2008/04/005",
      eprint         = "0802.1188",
      archivePrefix  = "arXiv",
      primaryClass   = "hep-ph",
      reportNumber   = "LPTHE-07-02",
      SLACcitation   = "%%CITATION = ARXIV:0802.1188;%%"
}

@article{Abbott:2018ikt,
      author         = "Abbott, B. and others",
      title          = "{Production and integration of the ATLAS Insertable
                        B-Layer}",
      collaboration  = "ATLAS IBL",
      journal        = "JINST",
      volume         = "13",
      year           = "2018",
      number         = "05",
      pages          = "T05008",
      doi            = "10.1088/1748-0221/13/05/T05008",
      eprint         = "1803.00844",
      archivePrefix  = "arXiv",
      primaryClass   = "physics.ins-det",
      SLACcitation   = "%%CITATION = ARXIV:1803.00844;%%"
}

@article{Czakon:2011xx:mod,
  author         = "Czakon, Michal and Mitov, Alexander",
  title          = "{Top++: A program for the calculation of the top-pair
                    cross-section at hadron colliders}",
  journal        = "Comput. Phys. Commun.",
  volume         = "185",
  year           = "2014",
  pages          = "2930",
  doi            = "10.1016/j.cpc.2014.06.021",
  eprint         = "1112.5675",
  archivePrefix  = "arXiv",
  primaryClass   = "hep-ph",
  reportNumber   = "CERN-PH-TH-2011-303, TTK-11-58",
}

@article{Aguilar-Saavedra:2014kpa,
      author         = "Aguilar-Saavedra, J. A. and Amidei, D. and Juste, A. and
                        Perez-Victoria, M.",
      title          = "{Asymmetries in top quark pair production at hadron
                        colliders}",
      journal        = "Rev. Mod. Phys.",
      volume         = "87",
      year           = "2015",
      pages          = "421-455",
      doi            = "10.1103/RevModPhys.87.421",
      eprint         = "1406.1798",
      archivePrefix  = "arXiv",
      primaryClass   = "hep-ph",
      reportNumber   = "CERN-PH-TH-2014-101",
      SLACcitation   = "%%CITATION = ARXIV:1406.1798;%%"
}

@article{Aguilar-Saavedra:2014vta,
    author = "Aguilar-Saavedra, J. A. and \'Alvarez, E. and Juste, A. and Rubbo, F.",
    title = "{Shedding light on the $t \bar t$ asymmetry: the photon handle}",
    eprint = "1402.3598",
    archivePrefix = "arXiv",
    primaryClass = "hep-ph",
    doi = "10.1007/JHEP04(2014)188",
    journal = "JHEP",
    volume = "04",
    pages = "188",
    year = "2014"
}

@article{Jung:2011zv,
      author         = "Jung, Sunghoon and Pierce, Aaron and Wells, James D.",
      title          = "{Top quark asymmetry from a non-Abelian horizontal
                        symmetry}",
      journal        = "Phys. Rev. D",
      volume         = "83",
      year           = "2011",
      pages          = "114039",
      doi            = "10.1103/PhysRevD.83.114039",
      eprint         = "1103.4835",
      archivePrefix  = "arXiv",
      primaryClass   = "hep-ph",
      SLACcitation   = "%%CITATION = ARXIV:1103.4835;%%"
}

@article{Diener:2009ee,
      author         = "Diener, Ross and Godfrey, Stephen and Martin, Travis A.
                        W.",
      title          = "{Using final state pseudorapidities to improve s-channel
                        resonance observables at the LHC}",
      journal        = "Phys. Rev. D",
      volume         = "80",
      year           = "2009",
      pages          = "075014",
      doi            = "10.1103/PhysRevD.80.075014",
      eprint         = "0909.2022",
      archivePrefix  = "arXiv",
      primaryClass   = "hep-ph",
      SLACcitation   = "%%CITATION = ARXIV:0909.2022;%%"
}

@article{Kidonakis:2010tc,
      author         = "Kidonakis, Nikolaos",
      title          = "{Next-to-next-to-leading logarithm resummation for s-channel single top quark
                        production}",
      journal        = "Phys. Rev. D",
      volume         = "81",
      year           = "2010",
      pages          = "054028",
      doi            = "10.1103/PhysRevD.81.054028",
      eprint         = "1001.5034",
      archivePrefix  = "arXiv",
      primaryClass   = "hep-ph",
      SLACcitation   = "%%CITATION = ARXIV:1001.5034;%%"
}

@article{LUCID2,
  author={G. Avoni and others},
  title={The new LUCID-2 detector for luminosity measurement and monitoring in ATLAS},
  journal={JINST},
  volume={13},
  number={07},
  pages={P07017},
  doi="10.1088/1748-0221/13/07/P07017",
  year={2018},
}

@article{Kidonakis:2011wy,
author         = "Kidonakis, Nikolaos",
title          = "{Next-to-next-to-leading-order collinear and soft gluon
                corrections for t-channel single top quark production}",
journal        = "Phys. Rev. D",
volume         = "83",
pages          = "091503",
doi            = "10.1103/PhysRevD.83.091503",
year           = "2011",
eprint         = "1103.2792",
archivePrefix  = "arXiv",
primaryClass   = "hep-ph",
SLACcitation   = "%%CITATION = ARXIV:1103.2792;%%",
}

@article{PhysRevD.86.034026,
    author = "Bernreuther, Werner and Si, Zong-Guo",
    title = "{Top quark and leptonic charge asymmetries for the Tevatron and LHC}",
    eprint = "1205.6580",
    archivePrefix = "arXiv",
    primaryClass = "hep-ph",
    reportNumber = "TTK-12-19",
    doi = "10.1103/PhysRevD.86.034026",
    journal = "Phys. Rev. D",
    volume = "86",
    pages = "034026",
    year = "2012"
}

@article{Aaltonen:2011kc,
      author         = "{CDF Collaboration}",
      title          = "{Evidence for a Mass Dependent Forward-Backward Asymmetry
                        in Top Quark Pair Production}",
      collaboration  = "CDF",
      journal        = "Phys. Rev. D",
      volume         = "83",
      year           = "2011",
      pages          = "112003",
      doi            = "10.1103/PhysRevD.83.112003",
      eprint         = "1101.0034",
      archivePrefix  = "arXiv",
      primaryClass   = "hep-ex",
      reportNumber   = "FERMILAB-PUB-10-525-E",
      SLACcitation   = "%%CITATION = ARXIV:1101.0034;%%"
}

@article{Aaltonen:2012it,
      author         = "{CDF Collaboration}",
      title          = "{Measurement of the top quark forward-backward production
                        asymmetry and its dependence on event kinematic
                        properties}",
      collaboration  = "CDF",
      journal        = "Phys. Rev. D",
      volume         = "87",
      year           = "2013",
      number         = "9",
      pages          = "092002",
      doi            = "10.1103/PhysRevD.87.092002",
      eprint         = "1211.1003",
      archivePrefix  = "arXiv",
      primaryClass   = "hep-ex",
      reportNumber   = "FERMILAB-PUB-12-591-E",
      SLACcitation   = "%%CITATION = ARXIV:1211.1003;%%"
}

@article{Abazov:2014cca,
      author         = "{D0 Collaboration}",
      title          = "{Measurement of the forward-backward asymmetry in top
                        quark-antiquark production in $p\bar{p}$ collisions using
                        the jepton+jets channel}",
      collaboration  = "D0",
      journal        = "Phys. Rev. D",
      volume         = "90",
      year           = "2014",
      pages          = "072011",
      doi            = "10.1103/PhysRevD.90.072011",
      eprint         = "1405.0421",
      archivePrefix  = "arXiv",
      primaryClass   = "hep-ex",
      reportNumber   = "FERMILAB-PUB-14-116-E",
      SLACcitation   = "%%CITATION = ARXIV:1405.0421;%%"
}

@article{Abazov:2015fna,
      author         = "{D0 Collaboration}",
      title          = "{Simultaneous measurement of forward-backward asymmetry
                        and top polarization in dilepton final states from $t\bar
                        t$ production at the Tevatron}",
      collaboration  = "D0",
      journal        = "Phys. Rev. D",
      volume         = "92",
      year           = "2015",
      pages          = "052007",
      doi            = "10.1103/PhysRevD.92.052007",
      eprint         = "1507.05666",
      archivePrefix  = "arXiv",
      primaryClass   = "hep-ex",
      reportNumber   = "FERMILAB-PUB-15-312-E",
      SLACcitation   = "%%CITATION = ARXIV:1507.05666;%%"
}

@article{PhysRevLett.115.052001,
    author = "Czakon, Michal and Fiedler, Paul and Mitov, Alexander",
    title = "{Resolving the Tevatron Top Quark Forward-Backward Asymmetry Puzzle: Fully Differential Next-to-Next-to-Leading-Order Calculation}",
    eprint = "1411.3007",
    archivePrefix = "arXiv",
    primaryClass = "hep-ph",
    reportNumber = "CAVENDISH-HEP-14-10, TTK-14-32",
    doi = "10.1103/PhysRevLett.115.052001",
    journal = "Phys. Rev. Lett.",
    volume = "115",
    number = "5",
    pages = "052001",
    year = "2015"
}

@article{PhysRevD.91.071502,
    author = "Kidonakis, Nikolaos",
    title = "{The top quark forward-backward asymmetry at approximate N$^3$LO}",
    eprint = "1501.01581",
    archivePrefix = "arXiv",
    primaryClass = "hep-ph",
    doi = "10.1103/PhysRevD.91.071502",
    journal = "Phys. Rev. D",
    volume = "91",
    number = "7",
    pages = "071502",
    year = "2015"
}

@article{arXiv:2208.12095,
     author         = "{ATLAS Collaboration}",
     title         = "{Evidence for the charge asymmetry in $pp \rightarrow
                       t\bar{t}$ production at $\sqrt{s}= 13$ TeV with the ATLAS
                       detector}",
      archivePrefix = "arXiv",
      eprint        = "2208.12095",
      reportNumber  = "CERN preprint ID: CERN-EP-2022-166",
      year          = "2022",
      url           = "https://cds.cern.ch/record/2825632",
}

@article{PhysRevD.98.014003,
    author = "Czakon, Michal and Heymes, David and Mitov, Alexander and Pagani, Davide and Tsinikos, Ioannis and Zaro, Marco",
    title = "{Top-quark charge asymmetry at the LHC and Tevatron through NNLO QCD and NLO EW}",
    eprint = "1711.03945",
    archivePrefix = "arXiv",
    primaryClass = "hep-ph",
    reportNumber = "CAVENDISH-HEP-17-13, CP3-17-46, NIKHEF-2017-59, TUM-HEP-1109-17, TTK-17-37",
    doi = "10.1103/PhysRevD.98.014003",
    journal = "Phys. Rev. D",
    volume = "98",
    number = "1",
    pages = "014003",
    year = "2018"
}

@misc{tensorflow2015-whitepaper,
title={ {TensorFlow}: Large-Scale Machine Learning on Heterogeneous Systems},
url={https://www.tensorflow.org/},
author={
    Abadi, Martin and
    Ashish~Agarwal and
    Paul~Barham and
    Eugene~Brevdo and
    Zhifeng~Chen and
    Craig~Citro and
    Greg~S.~Corrado and
    Andy~Davis and
    Jeffrey~Dean and
    Matthieu~Devin and
    Sanjay~Ghemawat and
    Ian~Goodfellow and
    Andrew~Harp and
    Geoffrey~Irving and
    Michael~Isard and
    Yangqing Jia and
    Rafal~Jozefowicz and
    Lukasz~Kaiser and
    Manjunath~Kudlur and
    Josh~Levenberg and
    Dandelion~Mane and
    Rajat~Monga and
    Sherry~Moore and
    Derek~Murray and
    Chris~Olah and
    Mike~Schuster and
    Jonathon~Shlens and
    Benoit~Steiner and
    Ilya~Sutskever and
    Kunal~Talwar and
    Paul~Tucker and
    Vincent~Vanhoucke and
    Vijay~Vasudevan and
    Fernanda~Viegas and
    Oriol~Vinyals and
    Pete~Warden and
    Martin~Wattenberg and
    Martin~Wicke and
    Yuan~Yu and
    Xiaoqiang~Zheng},
  year={2015},
}

@article{ATLAS:2022qxm,
  doi = {10.48550/ARXIV.2211.16345},
  author = {{ATLAS Collaboration}},
  title = {ATLAS flavour-tagging algorithms for the LHC Run 2 $pp$ collision dataset},
  publisher = {arXiv},
    eprint = "2211.16345",
    archivePrefix = "arXiv",
    primaryClass = "physics.data-an",
    reportNumber = "CERN-EP-2022-226",
    month = "11",
    year = "2022"
}

@article{ATLAS:2022swp,
    author = {{ATLAS Collaboration}},
    title = "{Tools for estimating fake/non-prompt lepton backgrounds with the ATLAS detector at the LHC}",
    eprint = "2211.16178",
    archivePrefix = "arXiv",
    primaryClass = "hep-ex",
    reportNumber = "CERN-EP-2022-214",
    month = "11",
    year = "2022"
}

@Article{Cacciari:2008gp,
     author    = "Cacciari, Matteo and Salam, Gavin P. and Soyez, Gregory",
     title     = "{The anti-\(k_{t}\) jet clustering algorithm}",
     journal   = "JHEP",
     volume    = "04",
     year      = "2008",
     pages     = "063",
     eprint    = "0802.1189",
     archivePrefix = "arXiv",
     primaryClass  =  "hep-ph",
     doi       = "10.1088/1126-6708/2008/04/063",
     SLACcitation  = "%%CITATION = 0802.1189;%%"
}

@Article{Fastjet,
      author         = "Cacciari, Matteo and Salam, Gavin P. and Soyez, Gregory",
      title          = "{FastJet user manual}",
      journal        = "Eur. Phys. J. C",
      volume         = "72",
      year           = "2012",
      pages          = "1896",
      doi            = "10.1140/epjc/s10052-012-1896-2",
      eprint         = "1111.6097",
      archivePrefix  = "arXiv",
      primaryClass   = "hep-ph",
      reportNumber   = "CERN-PH-TH-2011-297",
      SLACcitation   = "%%CITATION = ARXIV:1111.6097;%%"
}

@Article{Butterworth:2015oua,
      author         = "Butterworth, Jon and others",
      title          = "{PDF4LHC recommendations for LHC Run II}",
      journal        = "J. Phys. G",
      volume         = "43",
      year           = "2016",
      pages          = "023001",
      doi            = "10.1088/0954-3899/43/2/023001",
      eprint         = "1510.03865",
      archivePrefix  = "arXiv",
      primaryClass   = "hep-ph",
      reportNumber   = "OUTP-15-17P, SMU-HEP-15-12, TIF-UNIMI-2015-14,
                        LCTS-2015-27, CERN-PH-TH-2015-249",
      SLACcitation   = "%%CITATION = ARXIV:1510.03865;%%"
}

@Article{Pumplin:2002vw,
     author    = "Pumplin, J. and others",
     title     = "{New Generation of Parton Distributions with Uncertainties from Global QCD Analysis}",
     journal   = "JHEP",
     volume    = "07",
     year      = "2002",
     pages     = "012",
     doi       = "10.1088/1126-6708/2002/07/012",
     eprint    = "hep-ph/0201195",
     archivePrefix = "arXiv",
     SLACcitation  = "%%CITATION = HEP-PH/0201195;%%"
}

@article{Ball:2014uwa,
      author         = "{The NNPDF Collaboration} and Ball, Richard D. and others",
      title          = "{Parton distributions for the LHC run II}",
      journal        = "JHEP",
      volume         = "04",
      year           = "2015",
      pages          = "040",
      doi            = "10.1007/JHEP04(2015)040",
      eprint         = "1410.8849",
      archivePrefix  = "arXiv",
      primaryClass   = "hep-ph",
      reportNumber   = "EDINBURGH-2014-15, IFUM-1034-FT, CERN-PH-TH-2013-253,
                        OUTP-14-11P, CAVENDISH-HEP-14-11",
      SLACcitation   = "%%CITATION = ARXIV:1410.8849;%%"
}

@article{Melnikov:2011ta,
      author         = "Melnikov, Kirill and Schulze, Markus and Scharf, Andreas",
      title          = "{QCD corrections to top quark pair production in
                        association with a photon at hadron colliders}",
      journal        = "Phys. Rev. D",
      volume         = "83",
      year           = "2011",
      pages          = "074013",
      doi            = "10.1103/PhysRevD.83.074013",
      eprint         = "1102.1967",
      archivePrefix  = "arXiv",
      primaryClass   = "hep-ph",
      SLACcitation   = "%%CITATION = ARXIV:1102.1967;%%"
}

@Article{Sjostrand:2007gs,
	author    = "Sj{\"o}strand, T. and Mrenna, S. and Skands, P.",
	title     = "{A brief introduction to PYTHIA 8.1}",
	journal   = "Comput. Phys. Commun.",
	volume    = "178",
	year      = "2008",
	pages     = "852-867",
	eprint    = "0710.3820",
	archivePrefix = "arXiv",
	primaryClass  =  "hep-ph",
	doi       = "10.1016/j.cpc.2008.01.036",
	SLACcitation  = "%%CITATION = 0710.3820;%%"
}

@article{Sjostrand:2014zea,
      author         = "Sj{\"o}strand, Torbj{\"o}rn and Ask, Stefan and Christiansen,
                        Jesper R. and Corke, Richard and Desai, Nishita and Ilten,
                        Philip and Mrenna, Stephen and Prestel, Stefan and
                        Rasmussen, Christine O. and Skands, Peter Z.",
      title          = "{An introduction to PYTHIA 8.2}",
      journal        = "Comput. Phys. Commun.",
      volume         = "191",
      year           = "2015",
      pages          = "159",
      doi            = "10.1016/j.cpc.2015.01.024",
      eprint         = "1410.3012",
      archivePrefix  = "arXiv",
      primaryClass   = "hep-ph",
      reportNumber   = "LU-TP-14-36, MCNET-14-22, CERN-PH-TH-2014-190,
                        FERMILAB-PUB-14-316-CD, DESY-14-178, SLAC-PUB-16122,
                        --FERMILAB-PUB-14-316-CD",
      SLACcitation   = "%%CITATION = ARXIV:1410.3012;%%"
}

@Article{Lange:2001uf,
      author         = "Lange, D. J.",
      title          = "{The EvtGen particle decay simulation package}",
      booktitle      = "{Proceedings, 7th International Conference on B physics
                        at hadron machines (BEAUTY 2000)}",
      journal        = "Nucl. Instrum. Meth. A",
      volume         = "462",
      year           = "2001",
      pages          = "152",
      doi            = "10.1016/S0168-9002(01)00089-4",
      SLACcitation   = "%%CITATION = NUIMA,A462,152;%%"
}

@Article{Alwall:2014hca,
      author         = "Alwall, J. and Frederix, R. and Frixione, S. and Hirschi,
                        V. and Maltoni, F. and Mattelaer, O. and Shao, H. -S. and
                        Stelzer, T. and Torrielli, P. and Zaro, M.",
      title          = "{The automated computation of tree-level and
                        next-to-leading order differential cross sections, and
                        their matching to parton shower simulations}",
      journal        = "JHEP",
      volume         = "07",
      year           = "2014",
      pages          = "079",
      doi            = "10.1007/JHEP07(2014)079",
      eprint         = "1405.0301",
      archivePrefix  = "arXiv",
      primaryClass   = "hep-ph",
      reportNumber   = "CERN-PH-TH-2014-064, CP3-14-18, LPN14-066, MCNET-14-09,
                        ZU-TH-14-14",
      SLACcitation   = "%%CITATION = ARXIV:1405.0301;%%"
}

@article{Artoisenet:2012st,
      author         = "Artoisenet, Pierre and Frederix, Rikkert and Mattelaer,
                        Olivier and Rietkerk, Robbert",
      title          = "{Automatic spin-entangled decays of heavy resonances in
                        Monte Carlo simulations}",
      journal        = "JHEP",
      volume         = "03",
      pages          = "015",
      doi            = "10.1007/JHEP03(2013)015",
      year           = "2013",
      eprint         = "1212.3460",
      archivePrefix  = "arXiv",
      primaryClass   = "hep-ph",
      reportNumber   = "NIKHEF-2012-021, CERN-PH-TH-2012-329",
      SLACcitation   = "%%CITATION = ARXIV:1212.3460;%%",
}

@article{Frixione:2007zp,
      author         = "Frixione, Stefano and Laenen, Eric and Motylinski,
                        Patrick and Webber, Bryan R.",
      title          = "{Angular correlations of lepton pairs from vector boson
                        and top quark decays in Monte Carlo simulations}",
      journal        = "JHEP",
      volume         = "04",
      year           = "2007",
      pages          = "081",
      doi            = "10.1088/1126-6708/2007/04/081",
      eprint         = "hep-ph/0702198",
      archivePrefix  = "arXiv",
      OPTprimaryClass   = "HEP-PH",
      reportNumber   = "CAVENDISH-HEP-07-01, GEF-TH-09-2007, ITP-UU-07-10,
                        NIKHEF-2007-004",
      SLACcitation   = "%%CITATION = HEP-PH/0702198;%%"
}

@article{Frixione:1998jh,
      author         = "Frixione, Stefano",
      title          = "{Isolated photons in perturbative QCD}",
      journal        = "Phys. Lett. B",
      volume         = "429",
      year           = "1998",
      pages          = "369-374",
      doi            = "10.1016/S0370-2693(98)00454-7",
      eprint         = "hep-ph/9801442",
      archivePrefix  = "arXiv",
      XprimaryClass   = "hep-ph",
      reportNumber   = "ETH-TH-97-40",
      SLACcitation   = "%%CITATION = HEP-PH/9801442;%%"
}

@Article{Bahr:2008pv,
      author         = "B{\"a}hr, M. and others",
      title          = "{Herwig++ physics and manual}",
      journal        = "Eur. Phys. J. C",
      volume         = "58",
      year           = "2008",
      pages          = "639",
      doi            = "10.1140/epjc/s10052-008-0798-9",
      eprint         = "0803.0883",
      archivePrefix  = "arXiv",
      primaryClass   = "hep-ph",
      reportNumber   = "CERN-PH-TH-2008-038, CAVENDISH-HEP-08-03, KA-TP-05-2008,
                        DCPT-08-22, IPPP-08-11, CP3-08-05",
      SLACcitation   = "%%CITATION = ARXIV:0803.0883;%%"
}

@Article{Bellm:2015jjp,
      author         = "Bellm, Johannes and others",
      title          = "{Herwig 7.0/Herwig++ 3.0 release note}",
      journal        = "Eur. Phys. J. C",
      volume         = "76",
      year           = "2016",
      number         = "4",
      pages          = "196",
      doi            = "10.1140/epjc/s10052-016-4018-8",
      eprint         = "1512.01178",
      archivePrefix  = "arXiv",
      primaryClass   = "hep-ph",
      reportNumber   = "CERN-PH-TH-2015-289, MAN-HEP-2015-15, IFJPAN-IV-2015-13,
                        HERWIG-2015-01, KA-TP-18-2015, DCPT-15-142, MCNET-15-28,
                        IPPP-15-71, --HERWIG-2015-01",
      SLACcitation   = "%%CITATION = ARXIV:1512.01178;%%"
}

@Article{Campbell:1999ah,
      author         = "Campbell, John M. and Ellis, R. Keith",
      title          = "{Update on vector boson pair production at hadron colliders}",
      journal        = "Phys. Rev. D",
      volume         = "60",
      pages          = "113006",
      doi            = "10.1103/PhysRevD.60.113006",
      year           = "1999",
      eprint         = "hep-ph/9905386",
      archivePrefix  = "arXiv",
      reportNumber   = "FERMILAB-PUB-99-146-T",
      SLACcitation   = "%%CITATION = HEP-PH/9905386;%%",
}

@Article{Nason:2004rx,
      author         = "Nason, Paolo",
      title          = "{A new method for combining NLO QCD with shower Monte Carlo algorithms}",
      journal        = "JHEP",
      volume         = "11",
      pages          = "040",
      doi            = "10.1088/1126-6708/2004/11/040",
      year           = "2004",
      eprint         = "hep-ph/0409146",
      archivePrefix  = "arXiv",
}

@Article{Frixione:2007vw,
      author         = "Frixione, Stefano and Nason, Paolo and Oleari, Carlo",
      title          = "{Matching NLO QCD computations with parton shower
                        simulations: the POWHEG method}",
      journal        = "JHEP",
      volume         = "11",
      pages          = "070",
      doi            = "10.1088/1126-6708/2007/11/070",
      year           = "2007",
      eprint         = "0709.2092",
      archivePrefix  = "arXiv",
      primaryClass   = "hep-ph",
}

@Article{Alioli:2010xd,
      author         = "Alioli, Simone and Nason, Paolo and Oleari, Carlo and Re,
                        Emanuele",
      title          = "{A general framework for implementing NLO calculations in
                        shower Monte Carlo programs: the POWHEG BOX}",
      journal        = "JHEP",
      volume         = "06",
      pages          = "043",
      doi            = "10.1007/JHEP06(2010)043",
      year           = "2010",
      eprint         = "1002.2581",
      archivePrefix  = "arXiv",
      primaryClass   = "hep-ph",
}

@article{Kidonakis:2010ux,
      author         = "Kidonakis, Nikolaos",
      title          = "{Two-loop soft anomalous dimensions for single top quark
                        associated production with a \(W^{-}\) or \(H^{-}\)}",
      journal        = "Phys. Rev. D",
      volume         = "82",
      year           = "2010",
      pages          = "054018",
      doi            = "10.1103/PhysRevD.82.054018",
      eprint         = "1005.4451",
      archivePrefix  = "arXiv",
      primaryClass   = "hep-ph",
      SLACcitation   = "%%CITATION = ARXIV:1005.4451;%%"
}

@article{deFlorian:2016spz,
      author         = "de Florian, D. and others",
      title          = "{Handbook of LHC Higgs Cross Sections: 4. Deciphering the
                        Nature of the Higgs Sector}",
      collaboration  = "LHC Higgs Cross Section Working Group",
      doi            = "10.23731/CYRM-2017-002",
      year           = "2016",
      eprint         = "1610.07922",
      archivePrefix  = "arXiv",
      primaryClass   = "hep-ph",
      reportNumber   = "FERMILAB-FN-1025-T, CERN-2017-002-M",
      SLACcitation   = "%%CITATION = ARXIV:1610.07922;%%"
}

@Article{Hoeche:2009rj,
      author         = "H{\"o}che, Stefan and Krauss, Frank and Schumann, Steffen and Siegert, Frank",
      title          = "{QCD matrix elements and truncated showers}",
      journal        = "JHEP",
      volume         = "05",
      pages          = "053",
      doi            = "10.1088/1126-6708/2009/05/053",
      year           = "2009",
      eprint         = "0903.1219",
      archivePrefix  = "arXiv",
      primaryClass   = "hep-ph",
}

@article{Bothmann:2019yzt,
      author = "Bothmann, Enrico and others",
      title = "{Event generation with Sherpa 2.2}",
      journal = "SciPost Phys.",
      volume = "7",
      year = "2019",
      number = "3",
      pages = "034",
      doi = "10.21468/SciPostPhys.7.3.034",
      reportNumber = "FERMILAB-PUB-19-218-T, SLAC-PUB-17433, IPPP/19/42, MCNET-19-11",
      eprint = "1905.09127",
      archivePrefix = "arXiv",
      primaryClass = "hep-ph",
}

@article{Agostinelli:2002hh,
      author         = "Agostinelli, S. and others",
      title          = "{\textsc{Geant4} -- a simulation toolkit}",
      journal        = "Nucl. Instrum. Meth. A",
      volume         = "506",
      year           = "2003",
      pages          = "250",
      doi            = "10.1016/S0168-9002(03)01368-8",
      reportNumber   = "SLAC-PUB-9350, FERMILAB-PUB-03-339",
      SLACcitation   = "%%CITATION = NUIMA,A506,250;%%"
}

@online{chollet2015keras,
      author         = {Chollet, Fran\c{c}ois and others},
      title          = {Keras},
      year           = {2015},
      url            = {https://keras.io},
}

@Article{PERF-2007-01,
    author         = "{ATLAS Collaboration}",
    title          = "{The ATLAS Experiment at the CERN Large Hadron Collider}",
    journal        = "JINST",
    volume         = "3",
    year           = "2008",
    pages          = "S08003",
    doi            = "10.1088/1748-0221/3/08/S08003",
    primaryClass   = "hep-ex",
}

@Article{SOFT-2010-01,
    author         = "{ATLAS Collaboration}",
    title          = "{The ATLAS Simulation Infrastructure}",
    journal        = "Eur. Phys. J. C",
    volume         = "70",
    year           = "2010",
    pages          = "823",
    doi            = "10.1140/epjc/s10052-010-1429-9",
    eprint         = "1005.4568",
    archivePrefix  = "arXiv",
    primaryClass   = "physics.ins-det",
}

@Article{TOPQ-2012-17,
    author         = "{ATLAS Collaboration}",
    title          = "{Measurement of the top quark pair production charge asymmetry in proton--proton collisions at \(\sqrt{s} = 7\,\text{TeV}\) using the ATLAS detector}",
    journal        = "JHEP",
    volume         = "02",
    year           = "2014",
    pages          = "107",
    doi            = "10.1007/JHEP02(2014)107",
    reportNumber   = "CERN-PH-EP-2013-177",
    eprint         = "1311.6724",
    archivePrefix  = "arXiv",
    primaryClass   = "hep-ex",
}

@Article{TOPQ-2013-09,
    author         = "{ATLAS Collaboration}",
    title          = "{Measurement of the charge asymmetry in dileptonic decays of top quark pairs in \(pp\) collisions at \(\sqrt{s} = 7\,\text{TeV}\) using the ATLAS detector}",
    journal        = "JHEP",
    volume         = "05",
    year           = "2015",
    pages          = "061",
    doi            = "10.1007/JHEP05(2015)061",
    reportNumber   = "CERN-PH-EP-2014-233",
    eprint         = "1501.07383",
    archivePrefix  = "arXiv",
    primaryClass   = "hep-ex",
}

@Article{TOPQ-2014-16,
    author         = "{ATLAS Collaboration}",
    title          = "{Measurement of the charge asymmetry in top-quark pair production in the lepton-plus-jets final state in \(pp\) collision data at \(\sqrt{s} = 8\,\text{TeV}\) with the ATLAS detector}",
    journal        = "Eur. Phys. J. C",
    volume         = "76",
    year           = "2016",
    pages          = "87",
    doi            = "10.1140/epjc/s10052-016-3910-6",
    reportNumber   = "CERN-PH-EP-2015-217",
    eprint         = "1509.02358",
    archivePrefix  = "arXiv",
    primaryClass   = "hep-ex",
    related        = "TOPQ-2014-16-err",
    relatedstring  = "Erratum:",
}

@Article{PERF-2015-09,
    author         = "{ATLAS Collaboration}",
    title          = "{Jet reconstruction and performance using particle flow with the ATLAS Detector}",
    journal        = "Eur. Phys. J. C",
    volume         = "77",
    year           = "2017",
    pages          = "466",
    doi            = "10.1140/epjc/s10052-017-5031-2",
    reportNumber   = "CERN-EP-2017-024",
    eprint         = "1703.10485",
    archivePrefix  = "arXiv",
    primaryClass   = "hep-ex",
}

@Article{PERF-2015-10,
    author         = "{ATLAS Collaboration}",
    title          = "{Muon reconstruction performance of the ATLAS detector in proton--proton collision data at \(\sqrt{s} = 13\,\text{TeV}\)}",
    journal        = "Eur. Phys. J. C",
    volume         = "76",
    year           = "2016",
    pages          = "292",
    doi            = "10.1140/epjc/s10052-016-4120-y",
    reportNumber   = "CERN-EP-2016-033",
    eprint         = "1603.05598",
    archivePrefix  = "arXiv",
    primaryClass   = "hep-ex",
}

@Article{STDM-2015-03,
    author         = "{ATLAS Collaboration}",
    title          = "{Measurement of \(W^{\pm}\) and \(Z\)-boson production cross sections in \(pp\) collisions at \(\sqrt{s} = 13\,\text{TeV}\) with the ATLAS detector}",
    journal        = "Phys. Lett. B",
    volume         = "759",
    year           = "2016",
    pages          = "601",
    doi            = "10.1016/j.physletb.2016.06.023",
    reportNumber   = "CERN-EP-2016-069",
    eprint         = "1603.09222",
    archivePrefix  = "arXiv",
    primaryClass   = "hep-ex",
}

@Article{TOPQ-2015-08,
    author         = "{ATLAS Collaboration}",
    title          = "{Measurements of the charge asymmetry in top-quark pair production in the dilepton final state at \(\sqrt{s} = 8\,\text{TeV}\) with the ATLAS detector}",
    journal        = "Phys. Rev. D",
    volume         = "94",
    year           = "2016",
    pages          = "032006",
    doi            = "10.1103/PhysRevD.94.032006",
    reportNumber   = "CERN-EP-2016-051",
    eprint         = "1604.05538",
    archivePrefix  = "arXiv",
    primaryClass   = "hep-ex",
}

@Article{TOPQ-2015-10,
    author         = "{ATLAS Collaboration}",
    title          = "{Measurement of the charge asymmetry in highly boosted top-quark pair production in \(\sqrt{s} = 8\,\text{TeV}\) \(pp\) collision data collected by the ATLAS experiment}",
    journal        = "Phys. Lett. B",
    volume         = "756",
    year           = "2016",
    pages          = "52",
    doi            = "10.1016/j.physletb.2016.02.055",
    reportNumber   = "CERN-PH-EP-2015-303",
    eprint         = "1512.06092",
    archivePrefix  = "arXiv",
    primaryClass   = "hep-ex",
}

@Article{PERF-2016-04,
    author         = "{ATLAS Collaboration}",
    title          = "{Jet energy scale measurements and their systematic uncertainties in proton--proton collisions at \(\sqrt{s} = 13\,\text{TeV}\) with the ATLAS detector}",
    journal        = "Phys. Rev. D",
    volume         = "96",
    year           = "2017",
    pages          = "072002",
    doi            = "10.1103/PhysRevD.96.072002",
    reportNumber   = "CERN-EP-2017-038",
    eprint         = "1703.09665",
    archivePrefix  = "arXiv",
    primaryClass   = "hep-ex",
}

@Article{PERF-2016-05,
    author         = "{ATLAS Collaboration}",
    title          = "{Measurements of \(b\)-jet tagging efficiency with the ATLAS detector using \(t\bar{t}\) events at \(\sqrt{s} = 13\,\text{TeV}\)}",
    journal        = "JHEP",
    volume         = "08",
    year           = "2018",
    pages          = "089",
    doi            = "10.1007/JHEP08(2018)089",
    reportNumber   = "CERN-EP-2018-047",
    eprint         = "1805.01845",
    archivePrefix  = "arXiv",
    primaryClass   = "hep-ex",
}

@Article{PERF-2016-07,
    author         = "{ATLAS Collaboration}",
    title          = "{Performance of missing transverse momentum reconstruction with the ATLAS detector using proton--proton collisions at \(\sqrt{s} = 13\,\text{TeV}\)}",
    journal        = "Eur. Phys. J. C",
    volume         = "78",
    year           = "2018",
    pages          = "903",
    doi            = "10.1140/epjc/s10052-018-6288-9",
    reportNumber   = "CERN-EP-2017-274",
    eprint         = "1802.08168",
    archivePrefix  = "arXiv",
    primaryClass   = "hep-ex",
}

@Article{TRIG-2016-01,
    author         = "{ATLAS Collaboration}",
    title          = "{Performance of the ATLAS trigger system in 2015}",
    journal        = "Eur. Phys. J. C",
    volume         = "77",
    year           = "2017",
    pages          = "317",
    doi            = "10.1140/epjc/s10052-017-4852-3",
    reportNumber   = "CERN-EP-2016-241",
    eprint         = "1611.09661",
    archivePrefix  = "arXiv",
    primaryClass   = "hep-ex",
}

@Article{PERF-2017-02,
    author         = "{ATLAS Collaboration}",
    title          = "{Measurement of the photon identification efficiencies with the ATLAS detector using LHC Run~2 data collected in 2015 and 2016}",
    journal        = "Eur. Phys. J. C",
    volume         = "79",
    year           = "2019",
    pages          = "205",
    doi            = "10.1140/epjc/s10052-019-6650-6",
    reportNumber   = "CERN-EP-2018-216",
    eprint         = "1810.05087",
    archivePrefix  = "arXiv",
    primaryClass   = "hep-ex",
}

@Article{PERF-2017-03,
    author         = "{ATLAS Collaboration}",
    title          = "{Electron and photon energy calibration with the ATLAS detector using 2015--2016 LHC proton--proton collision data}",
    journal        = "JINST",
    volume         = "14",
    year           = "2019",
    pages          = "P03017",
    doi            = "10.1088/1748-0221/14/03/P03017",
    reportNumber   = "CERN-EP-2018-296",
    eprint         = "1812.03848",
    archivePrefix  = "arXiv",
    primaryClass   = "hep-ex",
}

@Article{TOPQ-2017-14,
    author         = "{ATLAS Collaboration}",
    title          = "{Measurements of inclusive and differential fiducial cross-sections of \(t\bar{t}\gamma\) production in leptonic final states at \(\sqrt{s} =  13\,\text{TeV}\) in ATLAS}",
    journal        = "Eur. Phys. J. C",
    volume         = "79",
    year           = "2019",
    pages          = "382",
    doi            = "10.1140/epjc/s10052-019-6849-6",
    reportNumber   = "CERN-EP-2018-302",
    eprint         = "1812.01697",
    archivePrefix  = "arXiv",
    primaryClass   = "hep-ex",
}

@Article{EGAM-2018-01,
    author         = "{ATLAS Collaboration}",
    title          = "{Electron and photon performance measurements with the ATLAS detector using the 2015--2017 LHC proton--proton collision data}",
    journal        = "JINST",
    volume         = "14",
    year           = "2019",
    pages          = "P12006",
    doi            = "10.1088/1748-0221/14/12/P12006",
    reportNumber   = "CERN-EP-2019-145",
    eprint         = "1908.00005",
    archivePrefix  = "arXiv",
    primaryClass   = "hep-ex",
}

@Article{FTAG-2018-01,
    author         = "{ATLAS Collaboration}",
    title          = "{ATLAS \(b\)-jet identification performance and efficiency measurement with \(t\bar{t}\) events in \(pp\) collisions at \(\sqrt{s} = 13\,\text{TeV}\)}",
    journal        = "Eur. Phys. J. C",
    volume         = "79",
    year           = "2019",
    pages          = "970",
    doi            = "10.1140/epjc/s10052-019-7450-8",
    reportNumber   = "CERN-EP-2019-132",
    eprint         = "1907.05120",
    archivePrefix  = "arXiv",
    primaryClass   = "hep-ex",
}

@Article{MUON-2018-03,
    author         = "{ATLAS Collaboration}",
    title          = "{Muon reconstruction and identification efficiency in ATLAS using the full Run~2 \(pp\) collision data set at \(\sqrt{s} = 13\,\text{TeV}\)}",
    journal        = "Eur. Phys. J. C",
    volume         = "81",
    year           = "2021",
    pages          = "578",
    doi            = "10.1140/epjc/s10052-021-09233-2",
    reportNumber   = "CERN-EP-2020-199",
    eprint         = "2012.00578",
    archivePrefix  = "arXiv",
    primaryClass   = "hep-ex",
}

@Article{TRIG-2018-01,
    author         = "{ATLAS Collaboration}",
    title          = "{Performance of the ATLAS muon triggers in Run~2}",
    journal        = "JINST",
    volume         = "15",
    year           = "2020",
    pages          = "P09015",
    doi            = "10.1088/1748-0221/15/09/p09015",
    reportNumber   = "CERN-EP-2020-031",
    eprint         = "2004.13447",
    archivePrefix  = "arXiv",
    primaryClass   = "hep-ex",
}

@Article{TRIG-2018-05,
    author         = "{ATLAS Collaboration}",
    title          = "{Performance of electron and photon triggers in ATLAS during LHC Run~2}",
    journal        = "Eur. Phys. J. C",
    volume         = "80",
    year           = "2020",
    pages          = "47",
    doi            = "10.1140/epjc/s10052-019-7500-2",
    reportNumber   = "CERN-EP-2019-169",
    eprint         = "1909.00761",
    archivePrefix  = "arXiv",
    primaryClass   = "hep-ex",
}

@Article{FTAG-2020-08,
    author         = "{ATLAS Collaboration}",
    title          = "{Measurement of the \(c\)-jet mistagging efficiency in \(t\bar{t}\) events using \(pp\) collision data at \(\sqrt{s} = 13\,\text{TeV}\) collected with the ATLAS detector}",
    journal        = "Eur. Phys. J. C",
    volume         = "82",
    year           = "2021",
    pages          = "95",
    doi            = "10.1140/epjc/s10052-021-09843-w",
    reportNumber   = "CERN-EP-2021-166",
    eprint         = "2109.10627",
    archivePrefix  = "arXiv",
    primaryClass   = "hep-ex",
}

@Article{TOPQ-2021-29,
    author         = "{ATLAS Collaboration}",
    title          = "{Measurement of the charge asymmetry in top-quark pair production in association with a photon with the ATLAS experiment}",
    year           = "2022",
    reportNumber   = "CERN-EP-2022-260",
    eprint         = "2212.10552",
    archivePrefix  = "arXiv",
    primaryClass   = "hep-ex",
}

@Booklet{ATL-SOFT-PUB-2021-001,
    author         = "{ATLAS Collaboration}",
    title          = "{The ATLAS Collaboration Software and Firmware}",
    howpublished   = "{ATL-SOFT-PUB-2021-001}",
    url            = "https://cds.cern.ch/record/2767187",
    year           = "2021",
}

@Booklet{ATL-SOFT-PUB-2021-003,
    author         = "{ATLAS Collaboration}",
    title          = "{ATLAS Computing Acknowledgements}",
    howpublished   = "{ATL-SOFT-PUB-2021-003}",
    url            = "https://cds.cern.ch/record/2776662",
    year           = "2021",
}

@Report{ATLAS-TDR-19,
    author         = "{ATLAS Collaboration}",
    title          = "{ATLAS Insertable B-Layer: Technical Design Report}",
    type           = "ATLAS-TDR-19; CERN-LHCC-2010-013",
    year           = "2010",
    url            = "https://cds.cern.ch/record/1291633",
    related        = "ATLAS-TDR-19-addm",
    relatedstring  = "Addendum:",
}

@Article{CMS-TOP-11-030,
    author         = "{CMS Collaboration}",
    title          = "{Inclusive and differential measurements of the \(t \bar{t}\) charge asymmetry in proton--proton collisions at \(\sqrt{s} = 7\,\text{TeV}\)}",
    journal        = "Phys. Lett. B",
    volume         = "717",
    year           = "2012",
    pages          = "129",
    doi            = "10.1016/j.physletb.2012.09.028",
    reportNumber   = "CERN-PH-EP-2012-175",
    eprint         = "1207.0065",
    archivePrefix  = "arXiv",
    primaryClass   = "hep-ex",
}

@Article{CMS-TOP-12-010,
    author         = "{CMS Collaboration}",
    title          = "{Measurements of the \(t\bar{t}\) charge asymmetry using the dilepton decay channel in \(pp\) collisions at \(\sqrt{s} = 7\,\text{TeV}\)}",
    journal        = "JHEP",
    volume         = "04",
    year           = "2014",
    pages          = "191",
    doi            = "10.1007/JHEP04(2014)191",
    reportNumber   = "CERN-PH-EP-2014-010",
    eprint         = "1402.3803",
    archivePrefix  = "arXiv",
    primaryClass   = "hep-ex",
}

@Article{CMS-TOP-12-033,
    author         = "{CMS Collaboration}",
    title          = "{Inclusive and differential measurements of the \(t\bar{t}\) charge asymmetry in \(pp\) collisions at \(\sqrt{s} = 8\,\text{TeV}\)}",
    journal        = "Phys. Lett. B",
    volume         = "757",
    year           = "2016",
    pages          = "154",
    doi            = "10.1016/j.physletb.2016.03.060",
    reportNumber   = "CERN-PH-EP-2015-141",
    eprint         = "1507.03119",
    archivePrefix  = "arXiv",
    primaryClass   = "hep-ex",
}

@Article{CMS-TOP-13-013,
    author         = "{CMS Collaboration}",
    title          = "{Measurement of the charge asymmetry in top quark pair production in \(pp\) collisions at \(\sqrt{s} = 8\,\text{TeV}\) using a template method}",
    journal        = "Phys. Rev. D",
    volume         = "93",
    year           = "2016",
    pages          = "034014",
    doi            = "10.1103/PhysRevD.93.034014",
    reportNumber   = "CERN-PH-EP-2015-189",
    eprint         = "1508.03862",
    archivePrefix  = "arXiv",
    primaryClass   = "hep-ex",
}

@Article{CMS-TOP-15-009,
    author         = "{CMS Collaboration}",
    title          = "{Measurements of \(t\bar{t}\) charge asymmetry using dilepton final states in \(pp\) collisions at \(\sqrt{s} = 8\,\text{TeV}\)}",
    journal        = "Phys. Lett. B",
    volume         = "760",
    year           = "2016",
    pages          = "365",
    doi            = "10.1016/j.physletb.2016.07.006",
    reportNumber   = "CERN-EP-2016-059",
    eprint         = "1603.06221",
    archivePrefix  = "arXiv",
    primaryClass   = "hep-ex",
}

@Article{CMS-TOP-17-014,
    author         = "{CMS Collaboration}",
    title          = "{Measurements of \(t\bar{t}\) differential cross sections in proton--proton collisions at \(\sqrt{s} = 13\,\text{TeV}\) using events containing two leptons}",
    journal        = "JHEP",
    volume         = "02",
    year           = "2019",
    pages          = "149",
    doi            = "10.1007/JHEP02(2019)149",
    reportNumber   = "CERN-EP-2018-252",
    eprint         = "1811.06625",
    archivePrefix  = "arXiv",
    primaryClass   = "hep-ex",
}

@Booklet{ATLAS-CONF-2014-018,
    author         = "{ATLAS Collaboration}",
    title          = "{Tagging and suppression of pileup jets with the ATLAS detector}",
    howpublished   = "{ATLAS-CONF-2014-018}",
    url            = "https://cds.cern.ch/record/1700870",
    year           = "2014",
}

@Booklet{ATLAS-CONF-2018-006,
    author         = "{ATLAS Collaboration}",
    title          = "{Calibration of light-flavour \(b\)-jet mistagging rates using ATLAS proton--proton collision data at \(\sqrt{s} = 13~\text{TeV}\)}",
    howpublished   = "{ATLAS-CONF-2018-006}",
    url            = "https://cds.cern.ch/record/2314418",
    year           = "2018",
}

@Booklet{ATLAS-CONF-2018-023,
    author         = "{ATLAS Collaboration}",
    title          = "{\(E_{\text{T}}^{\text{miss}}\) performance in the ATLAS detector using 2015--2016 LHC \(pp\) collisions}",
    howpublished   = "{ATLAS-CONF-2018-023}",
    url            = "https://cds.cern.ch/record/2625233",
    year           = "2018",
}

@Booklet{ATLAS-CONF-2019-021,
    author         = "{ATLAS Collaboration}",
    title          = "{Luminosity determination in \(pp\) collisions at \(\sqrt{s} = 13\,\text{TeV}\) using the ATLAS detector at the LHC}",
    howpublished   = "{ATLAS-CONF-2019-021}",
    url            = "https://cds.cern.ch/record/2677054",
    year           = "2019",
}

@Booklet{ATL-PHYS-PUB-2010-013,
    author         = "{ATLAS Collaboration}",
    title          = "{The simulation principle and performance of the ATLAS fast calorimeter simulation FastCaloSim}",
    howpublished   = "{ATL-PHYS-PUB-2010-013}",
    url            = "https://cds.cern.ch/record/1300517",
    year           = "2010",
}

@Booklet{ATL-PHYS-PUB-2012-003,
    author         = "{ATLAS Collaboration}",
    title          = "{Summary of ATLAS Pythia 8 tunes}",
    howpublished   = "{ATL-PHYS-PUB-2012-003}",
    url            = "https://cds.cern.ch/record/1474107",
    year           = "2012",
}

@Booklet{ATL-PHYS-PUB-2014-021,
    author         = "{ATLAS Collaboration}",
    title          = "{ATLAS Pythia~8 tunes to \(7~\text{TeV}\) data}",
    howpublished   = "{ATL-PHYS-PUB-2014-021}",
    url            = "https://cds.cern.ch/record/1966419",
    year           = "2014",
}

@Booklet{ATL-PHYS-PUB-2015-015,
    author         = "{ATLAS Collaboration}",
    title          = "{Jet Calibration and Systematic Uncertainties for Jets Reconstructed in the ATLAS Detector at \(\sqrt{s} = 13~\text{TeV}\)}",
    howpublished   = "{ATL-PHYS-PUB-2015-015}",
    url            = "https://cds.cern.ch/record/2037613",
    year           = "2015",
}

@Booklet{ATL-PHYS-PUB-2016-017,
    author         = "{ATLAS Collaboration}",
    title          = "{The Pythia~8 A3 tune description of ATLAS minimum bias and inelastic measurements incorporating the Donnachie--Landshoff diffractive model}",
    howpublished   = "{ATL-PHYS-PUB-2016-017}",
    url            = "https://cds.cern.ch/record/2206965",
    year           = "2016",
}

@Booklet{ATL-PHYS-PUB-2018-009,
    author         = "{ATLAS Collaboration}",
    title          = "{Improvements in \(t\bar{t}\) modelling using NLO+PS Monte Carlo generators for Run~2}",
    howpublished   = "{ATL-PHYS-PUB-2018-009}",
    url            = "https://cds.cern.ch/record/2630327",
    year           = "2018",
}
 
\clearpage
 
\begin{flushleft}
\hypersetup{urlcolor=black}
{\Large The ATLAS Collaboration}

\bigskip

\AtlasOrcid[0000-0002-6665-4934]{G.~Aad}$^\textrm{\scriptsize 101}$,
\AtlasOrcid[0000-0002-5888-2734]{B.~Abbott}$^\textrm{\scriptsize 119}$,
\AtlasOrcid[0000-0002-7248-3203]{D.C.~Abbott}$^\textrm{\scriptsize 102}$,
\AtlasOrcid[0000-0002-1002-1652]{K.~Abeling}$^\textrm{\scriptsize 55}$,
\AtlasOrcid[0000-0002-8496-9294]{S.H.~Abidi}$^\textrm{\scriptsize 29}$,
\AtlasOrcid[0000-0002-9987-2292]{A.~Aboulhorma}$^\textrm{\scriptsize 35e}$,
\AtlasOrcid[0000-0001-5329-6640]{H.~Abramowicz}$^\textrm{\scriptsize 150}$,
\AtlasOrcid[0000-0002-1599-2896]{H.~Abreu}$^\textrm{\scriptsize 149}$,
\AtlasOrcid[0000-0003-0403-3697]{Y.~Abulaiti}$^\textrm{\scriptsize 116}$,
\AtlasOrcid[0000-0003-0762-7204]{A.C.~Abusleme~Hoffman}$^\textrm{\scriptsize 136a}$
% ...
\bigskip
\\

% ...
$^{29}$Physics Department, Brookhaven National Laboratory, Upton NY; United States of America.\\
% ...
$^{35}$$^{(a)}$Facult\'e des Sciences Ain Chock, R\'eseau Universitaire de Physique des Hautes Energies - Universit\'e Hassan II, Casablanca;$^{(b)}$Facult\'{e} des Sciences, Universit\'{e} Ibn-Tofail, K\'{e}nitra;$^{(c)}$Facult\'e des Sciences Semlalia, Universit\'e Cadi Ayyad, LPHEA-Marrakech;$^{(d)}$LPMR, Facult\'e des Sciences, Universit\'e Mohamed Premier, Oujda;$^{(e)}$Facult\'e des sciences, Universit\'e Mohammed V, Rabat;$^{(f)}$Institute of Applied Physics, Mohammed VI Polytechnic University, Ben Guerir; Morocco.\\
% ...
$^{55}$II. Physikalisches Institut, Georg-August-Universit\"{a}t G\"ottingen, G\"ottingen; Germany.\\
% ...
$^{101}$CPPM, Aix-Marseille Universit\'e, CNRS/IN2P3, Marseille; France.\\
$^{102}$Department of Physics, University of Massachusetts, Amherst MA; United States of America.\\
% ...
$^{116}$Department of Physics, New York University, New York NY; United States of America.\\
% ...
$^{119}$Homer L. Dodge Department of Physics and Astronomy, University of Oklahoma, Norman OK; United States of America.\\
% ...
$^{136}$$^{(a)}$Departamento de F\'isica, Pontificia Universidad Cat\'olica de Chile, Santiago;$^{(b)}$Millennium Institute for Subatomic physics at high energy frontier (SAPHIR), Santiago;$^{(c)}$Instituto de Investigaci\'on Multidisciplinario en Ciencia y Tecnolog\'ia, y Departamento de F\'isica, Universidad de La Serena;$^{(d)}$Universidad Andres Bello, Department of Physics, Santiago;$^{(e)}$Instituto de Alta Investigaci\'on, Universidad de Tarapac\'a, Arica;$^{(f)}$Departamento de F\'isica, Universidad T\'ecnica Federico Santa Mar\'ia, Valpara\'iso; Chile.\\
% ...
$^{149}$Department of Physics, Technion, Israel Institute of Technology, Haifa; Israel.\\
$^{150}$Raymond and Beverly Sackler School of Physics and Astronomy, Tel Aviv University, Tel Aviv; Israel.\\
% ...
$^{a}$ Also Affiliated with an institute covered by a cooperation agreement with CERN.\\
$^{b}$ Also at An-Najah National University, Nablus; Palestine.\\
$^{c}$ Also at Borough of Manhattan Community College, City University of New York, New York NY; United States of America.\\
$^{d}$ Also at Bruno Kessler Foundation, Trento; Italy.\\
$^{e}$ Also at Center for High Energy Physics, Peking University; China.\\
$^{f}$ Also at Center for Interdisciplinary Research and Innovation (CIRI-AUTH), Thessaloniki ; Greece.\\
% ...

\end{flushleft}


\end{document}